\documentclass[12pt]{article}
\usepackage{jheppub}
\pdfoutput=1

\usepackage{amsmath,bbm,array,amsfonts,graphicx,wrapfig,lscape,float,mathtools,slashbox,multirow,longtable}

\newcommand{\be}{\begin{equation}}
\newcommand{\ee}{\end{equation}}
\newcommand{\beq}{\begin{equation}}
\newcommand{\beql}[1]{\begin{equation}\label{#1}}
\newcommand{\eeq}{\end{equation}}
\newcommand{\ba}{\begin{array}}
\newcommand{\ea}{\end{array}}
\newcommand{\bea}{\begin{eqnarray}}
\newcommand{\beal}[1]{\begin{eqnarray}\label{#1}}
\newcommand{\eea}{\end{eqnarray}}
\newcommand{\ben}{\begin{enumerate}}
\newcommand{\een}{\end{enumerate}}
\newcommand{\bean}{\begin{eqnarray*}}
\newcommand{\eean}{\end{eqnarray*}}
\newcommand{\eref}[1]{(\ref{#1})}
\newcommand{\sref}[1]{\S\ref{#1}}
\newcommand{\tref}[1]{Table~\ref{#1}}
\newcommand{\nn}{\nonumber}

\newcommand{\fref}[1]{Figure \ref{#1}}
\newcommand{\btab}[1]{\begin{tabular}{#1}}
\newcommand{\etab}{\end{tabular}}

\newcommand{\master}{{}^{\text{Irr}}\mathcal{F}^\flat}
\newcommand{\mesonic}{\mathcal{M}^{mes}}

\def \Black {\pdfsetcolor{0 0 0 1}}

\newcommand{\comment}[1]{}

\newcommand{\ud}{\mathrm{d}}

\newcommand{\qed}{\nobreak \ifvmode \relax \else
      \ifdim\lastskip<1.5em \hskip-\lastskip
      \hskip1.5em plus0em minus0.5em \fi \nobreak
      \vrule height0.75em width0.5em depth0.25em\fi}

\title{Brane Tilings and Specular Duality}

\author{Amihay Hanany}
\author{and Rak-Kyeong Seong}

\affiliation{
Theoretical Physics Group, The Blackett Laboratory,
Imperial College London, \\
Prince Consort Road, London SW7 2AZ, UK
}

\emailAdd{a.hanany@imperial.ac.uk}
\emailAdd{rak-kyeong.seong@imperial.ac.uk}

\preprint{Imperial/TP/12/AH/03}

\abstract{
We study a new duality which pairs $4d$ $\mathcal{N}=1$ supersymmetric quiver gauge theories. They are represented by brane tilings and are worldvolume theories of D3 branes at Calabi-Yau 3-fold singularities. The new duality identifies theories which have the same combined mesonic and baryonic moduli space, otherwise called the master space. We obtain the associated Hilbert series which encodes both the generators and defining relations of the moduli space. We illustrate our findings with a set of brane tilings that have reflexive toric diagrams.
}

\begin{document}

\maketitle

\section{Introduction}

Dualities have vastly contributed towards a better understanding of string theory and beyond. A particular example is mirror symmetry \cite{Lerche:1989uy,Candelas:1989hd,Greene:1990ud,morrison-1993-6,Batyrev:1994hm,Batyrev:1994ju,Batyrev:1997tv,cox1999mirror,mirrorbook} which identifies two Type II superstring theories compactified on Calabi-Yau 3-folds whose Hodge numbers are swapped. A similar example, although only true at low energies, is \textit{toric (Seiberg) duality} \cite{Feng:2000mi,Feng:2001xr,Feng:2002zw,Seiberg:1994pq,Feng:2001bn,2001JHEP...12..001B,Franco:2003ea}. It relates supersymmetric worldvolume theories of D3-branes on singular toric Calabi-Yau 3-folds which have isomorphic mesonic moduli spaces.

\begin{figure}[ht!!]
\begin{center}
\resizebox{0.801\hsize}{!}{
\includegraphics[trim=0cm 0cm 0cm 0cm,totalheight=16 cm]{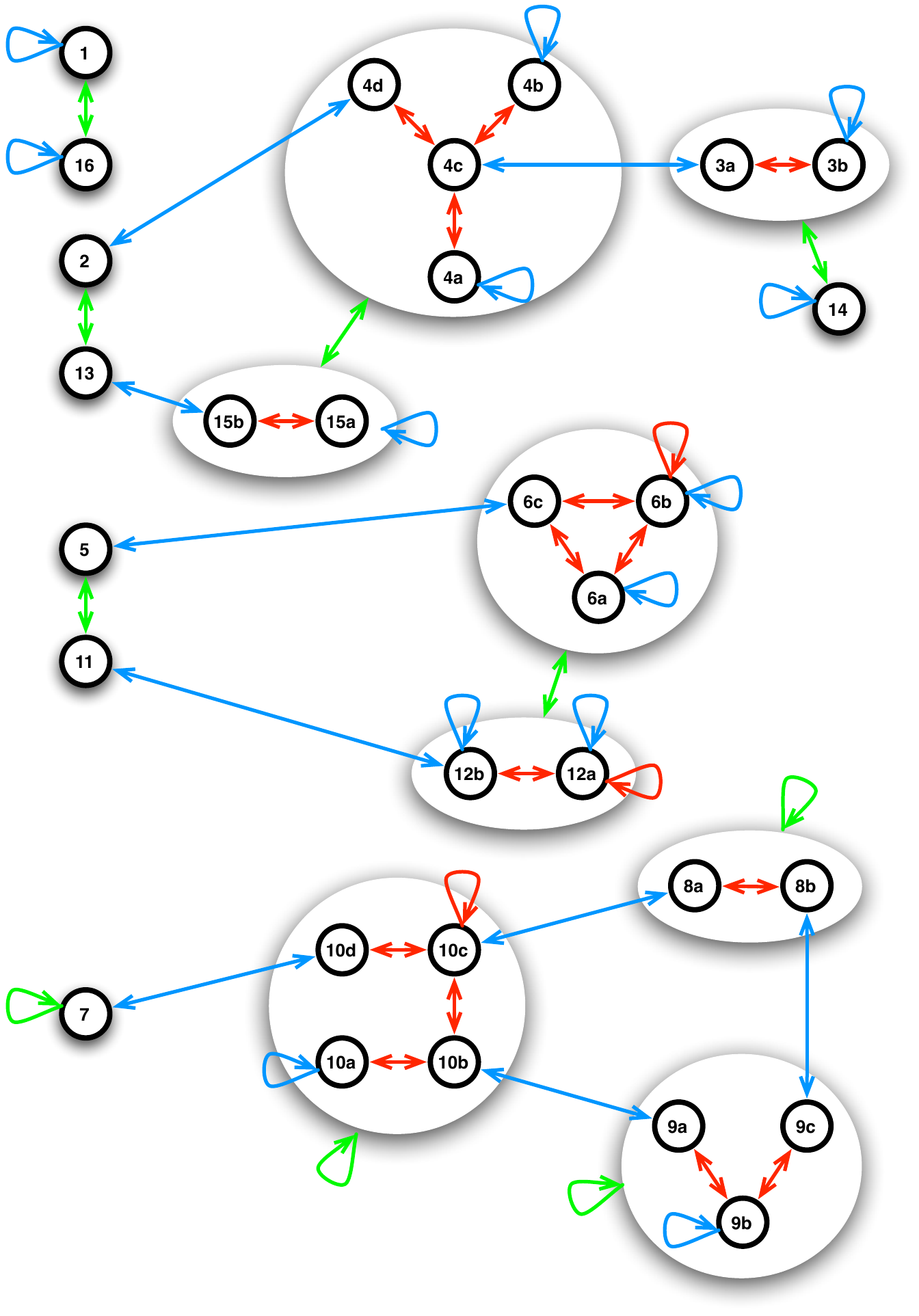}
}  \caption{\textit{The three dualities for Brane Tilings with Reflexive Toric Diagrams.} The arrows indicate toric duality (red), specular duality (blue), and reflexive duality (green) which is discussed in \cite{Hanany:2012hi}. The black nodes of the duality tree represent distinct brane tilings, where the labels are taken from \cite{Hanany:2012hi} and \fref{f_sumtoric2}.
  \label{ctree3}}
 \end{center}
 \end{figure}

These $4d$ $\mathcal{N}=1$ supersymmetric field theories have mesonic moduli spaces which are toric Calabi-Yau 3-folds. Their geometry is encoded in a convex lattice polygon called the toric diagram. Furthermore, the theories are best expressed by periodic bipartite graphs on $\mathbb{T}^2$, otherwise known as brane tilings \cite{Hanany:2005ve,Franco:2005rj,Franco:2005sm,Hanany:2005ss,Hanany:2006nm,Kennaway:2007tq,Yamazaki:2008bt}. They represent the largest known class of supersymmetric field theories that are associated to toric Calabi-Yau 3-folds. 

The rich combinatorial structure of brane tilings led recently to new insights beyond toric duality. For instance, certain toric diagrams have a single interior point and exhibit the special property of appearing in polar dual pairs \cite{1997CMaPh.185..495K,Kreuzer:1998vb,Kreuzer:2000qv,Kreuzer:2000xy,2008arXiv0802.3376B,2008arXiv0809.4681C}. They are called reflexive toric diagrams and relate to a correspondence between brane tilings which was studied in \cite{Hanany:2012hi}. Given brane tilings A and B whose reflexive toric diagrams are a dual pair, the toric diagram of brane tiling A is the lattice of generators of the mesonic moduli space of brane tiling B, and vice versa.  We call this correspondence \textit{reflexive duality}.

In the following, we discuss a new correspondence that was named in \cite{Hanany:2012hi} \textbf{specular duality}. It identifies brane tilings which have isomorphic combined mesonic and baryonic moduli spaces, also known as master spaces $\mathcal{F}^\flat$. The following scenarios of brane tilings apply to this new duality:
\begin{itemize}
\item[1.] Dual brane tilings A and B are both on $\mathbb{T}^2$. They have reflexive toric diagrams.

\item[2.] Brane tiling A is on $\mathbb{T}^2$ and dual brane tiling B is not on $\mathbb{T}^2$. Brane tiling A has a toric diagram which is not reflexive.

\item[3.] Both brane tilings A and B are not on $\mathbb{T}^2$.
\end{itemize}

For brane tilings with reflexive toric diagrams, specular duality manifests itself not only as an isomorphism between master spaces. The additional properties are:
\begin{itemize}
\item The external/internal perfect matchings of brane tiling A are the internal/external perfect matchings of brane tiling B.
\item The mesonic flavour symmetries of brane tiling A are the hidden or anomalous baryonic symmetries of brane tiling B, and vice versa.
\end{itemize} 

The following paper studies specular duality restricted to brane tilings with reflexive toric diagrams. The Hilbert series of $\mathcal{F}^\flat$ is computed explicitly to illustrate its invariance under the new correspondence. The swap between external and internal perfect matchings, and mesonic and baryonic symmetries is explained. Moreover, we illustrate that specular duality is a reflection of the Calabi-Yau cone of $\mathcal{F}^\flat$ along a hyperplane.

The new correspondence extends beyond brane tilings with reflexive toric diagrams. Accordingly, specular duality can lead to brane tilings on spheres or Riemann surfaces with genus $g\geq 2$. These have no known AdS dual and have mesonic moduli spaces which are not necessarily Calabi-Yau 3-folds \cite{Benvenuti:2004dw,Benvenuti:2005wi,Kennaway:2007tq}. Their quiver and superpotential however admit a master space which can be traced back to a brane tiling on $\mathbb{T}^2$.

Specular duality for brane tilings that are not on $\mathbb{T}^2$ may lead to new insights into quiver gauge theories and Calabi-Yau moduli spaces. The work concludes with this observation and highlights the importance of future studies \cite{HananySeong11b}.
\\

The paper is divided into the following sections. Section \sref{s2} reviews brane tilings and their mesonic moduli spaces and master spaces. They are analysed with the help of Hilbert series. Section \sref{s3} begins with a short review on toric duality and compares its properties with the characteristics of specular duality. The new correspondence between brane tilings is explained in terms of the untwisting map \cite{Feng:2005gw,Franco:2011sz,Stienstra:2007dy} and  modified shivers \cite{Butti:2007jv,Franco:2007ii,Hanany:2008fj}. Section \sref{s4} studies and summarises the transformation of the brane tiling, the exchange of perfect matchings, and the swap of mesonic and baryonic symmetries under specular duality. The concluding section gives a short introduction on how specular duality relates brane tilings on $\mathbb{T}^2$ with tilings on spheres and Riemann surfaces of genus $g\geq 2$. 
\\

\section{Brane Tilings and their Moduli Spaces \label{s2}}

The following section reviews brane tilings and their mesonic moduli spaces and master spaces. The method of calculating Hilbert series is reviewed. Readers who are familiar with these topics may skip the section and move on to the discussion of specular duality in Section \sref{s3}.
\\

\subsection{Brane Tilings, F- and D-term charges, and the Toric Diagram \label{s2a}}

A brane tiling represents a worldvolume theory of D3 branes on a singular non-compact Calabi-Yau 3-fold. It encodes the bifundamental matter content and superpotential of the theory.
\\

A \textbf{quiver} is a graph which encodes as a set of $G$ nodes the $U(N)_i$ gauge groups and as a set of $e$ arrows the bifundamental fields $X_{ij}$ of the gauge theory. The number of incoming and outgoing arrows is the same at each node. The incidence matrix $d_{G\times e}$ of the graph encodes this property with its vanishing sum of rows.\footnote{$d_{G\times e}$ can therefore be reduced to its $G-1$ independent rows $\Delta_{(G-1)\times e}$.} This property is called the quiver's Calabi-Yau condition \cite{Feng:2002zw,Forcella:2008bb,Forcella:2008eh}.

A \textbf{brane tiling} or dimer \cite{Hanany:2005ve,Franco:2005rj,Franco:2005sm,Kennaway:2007tq,2007arXiv0710.1898I} is a periodic bipartite graph on $\mathbb{T}^2$. It has the following components:
\begin{itemize}
	\item \textbf{Faces} relate to $U(N)_i$ gauge groups. The ranks $N$ of all gauge groups are the same and equal to the number of probe D3-branes. 
	\item \textbf{Edges} relate to the bifundamental fields. Every edge $X_{ij}$ in the tiling has two neighbouring faces $U(N)_i$ and $U(N)_j$. The orientation of the bifundamental field $X_{ij}$ is given by the orientation around the black and white nodes at the two ends of the corresponding edge.
	\item \textbf{White (resp. black) nodes} correspond to positive (negative) terms in the superpotential $W$. They have a clockwise (anti-clockwise) orientation. By following the orientation around a node, one can identify the fields of the related superpotential term in the correct cyclic order.
\end{itemize}

The geometry of the toric Calabi-Yau 3-fold is encoded in the brane tiling. A new basis of fields is defined from the set of quiver fields in order to describe both F-term and D-term constraints. The new fields are known as gauge linear sigma model (GLSM) fields \cite{Witten:1993yc} and are represented as perfect matchings \cite{Hanany:2005ve,Hanany:2005ss,Kennaway:2007tq,Forcella:2008bb} of the brane tiling:\begin{itemize}
	\item A \textbf{perfect matching} $p_\alpha$ is a set of bifundamental fields which connects to all nodes in the brane tiling precisely once. It corresponds to a point in the toric diagram of the Calabi-Yau 3-fold. A perfect matchings which relates to an \textbf{extremal (corner)} point of the toric diagram has non-zero IR $U(1)_R$ charge. An \textbf{internal} as well as a non-extremal toric point on the perimeter of the toric diagram has zero R-charge. We call all points on the perimeter \textbf{external}, including extremal ones. The number of internal, external and extremal perfect matchings is denoted by $n_i$, $n_e$ and $n_p$ respectively. All perfect matchings are summarized in a matrix $P_{e\times c}$ \cite{Forcella:2008bb}, where $e$ is the number of matter fields and $c$ the number of perfect matchings. The perfect matching matrix $P_{e\times c}$ takes the form
	\beal{es00_1cc00}
	P_{i\alpha}= \left\{
	\ba{ll} 1 & ~~\text{if} ~X_i \in p_\alpha \\ 
	0 & ~~\text{if} ~X_i \notin p_\alpha \ea \right.
	~~,
	\eea
	where $i=1,\dots,e$ and $\alpha=1,\dots,c$.
	\item \textbf{F-terms} $\partial_X W =0$ are encoded in $P_{e\times c}$. The charges under the F-term constraints are given by the kernel,
	\beal{es00_1c0}
	Q_{F~(c-G-2)\times c} =  \ker{(P_{e \times c})}~~,
	\eea
	where $G$ is the number of gauge groups.\footnote{Note: $\ker$ used here takes the transpose of the matrix.}
	\item \textbf{D-terms} are encoded in the quiver incidence matrix $d_{G\times e}$. The charges $Q_{D~(G-1)\times c}$ under the D-term constraints are defined by
	\beal{es00_3}
\Delta_{(G-1)\times E} = Q_{D~(G-1)\times c}.P^{t}_{c \times e}~~.
\eea
\end{itemize}

The F- and D-term charge matrices are concatenated to form a total charge matrix
\beal{es00_4}
Q_{t~(c-3)\times c} =  \left(\ba{c} Q_F \\ Q_D\ea\right)~~.
\eea
The kernel of $Q_t$,
\beal{es00_5}
G_t = \ker(Q_t)~~,
\eea
corresponds to a matrix whose columns relate to perfect matchings. The rows of $G_t$ are the coordinates of the associated point in the toric diagram.
\\

\subsection{The Master Space $\mathcal{F}^\flat$ and the Mesonic Moduli Space $\mathcal{M}^{mes}$ \label{s2b}}

\noindent\textbf{Master Space $\mathcal{F}^\flat$.} The master space is the combined mesonic and baryonic moduli space. It has the following properties:
\begin{itemize}
\item The \textbf{master space} \cite{Forcella:2008bb,Forcella:2008eh,Hanany:2010zz} of the one D3-brane theory relates to the following quotient ring
\beal{es2b_1}
\mathbb{C}^{e}[X_{1},\dots,X_{e}]/\mathcal{I}_{\partial W =0}~~,
\eea
where $e$ is the number of bifundamental fields $X_i$. $\mathbb{C}^{e}[X_{1},\dots,X_{e}]$ is the complex ring over all bifundamental fields, and $\mathcal{I}_{\partial W=0}$ is the ideal formed by the F-terms. 

\item The master space in \eref{es2b_1} is usually reducible into components. The largest irreducible component is known as the \textbf{coherent component} ${}^{\text{Irr}}\mathcal{F}^\flat$ and is toric Calabi-Yau. All other smaller components are generally linear pieces of the form $\mathbb{C}^l$. In our discussion, we will concentrate on the coherent component of the master space and for simplicity use $\mathcal{F}^\flat$ and $\master$ interchangeably.

\item The coherent component of the master space is the following \textbf{symplectic quotient}
\beal{es2b_2}
\master = \mathbb{C}^{c}[p_1,\dots,p_c]// Q_F ~~,
\eea
where $c$ is the number of perfect matchings $p_\alpha$ in the brane tiling. The symplectic quotient captures invariants of the ring $\mathbb{C}^c[p_1,\dots,p_c]$ under the charges $Q_F$ in \eref{es00_1c0}.

\item The \textbf{dimension} of the master space $\mathcal{F}^\flat$ is $G+2$. 
\comment{The mesonic symmetry is $U(1)^3$ or an enhancement of $U(1)^3$ with rank $3$ where one $U(1)$ is R and two $U(1)$'s are flavor symmetries. The baryonic symmetry is $U(1)^{G-1}$ or an enhancement of $U(1)^{G-1}$ with rank $G-1$ which includes both \textbf{anomalous} or enhanced \textbf{hidden} symmetries and \textbf{non-anomalous} baryonic symmetries.}

\end{itemize}

\noindent The master space exhibits the following symmetries:
\begin{itemize}
\item The \textbf{mesonic symmetry} is $U(1)^3$ or an enhancement with rank $3$. An enhancement is indicated by extremal perfect matchings which carry the same $Q_F$ charges. The mesonic symmetry contains the $U(1)_R$ symmetry and the flavor symmetries. It derives from the isometry of the toric Calabi-Yau 3-fold.
\item The \textbf{baryonic symmetry} is $U(1)^{G-1}$ or an enhancement with rank $G-1$. An enhancement is indicated by non-extremal perfect matchings which carry the same $Q_F$ charges. It contains both anomalous and non-anomalous symmetries which have decoupling gauge dynamics in the IR. Non-Abelian extensions of these symmetries are known as \textbf{hidden symmetries} \cite{Forcella:2008bb,Forcella:2008eh,Hanany:2010zz}.
\end{itemize}

\noindent Let $I$ and $E$ denote respectively the number of internal and external points in the toric diagram.\footnote{Note: Points in the toric diagram can carry multiplicities according to the number of perfect matchings associated to them. $I$ and $E$ is a counting that ignores multiplicities and hence there is no direct correspondence to the number of perfect matchings $n_i$, $n_e$ and $n_p$.} They are used to define the following quantities:
\begin{itemize}
\item The  number of \textbf{anomalous} $U(1)$ baryonic symmetries or the total rank of enhanced \textbf{hidden} baryonic symmetries is given by $2 I$.
\item The number of \textbf{non-anomalous} baryonic $U(1)$'s is $E -3$.
\item The total number of baryonic symmetries is as stated above $G-1$. Accordingly,
\beal{es2c_10}
G-1 = 2 I + E -3 ~\Rightarrow~ 
A=\frac{G}{2} = I + \frac{E}{2} - 1
\eea
which is \textbf{Pick's theorem} generalised to toric diagrams. The unit square area $A$ of a toric diagram is scaled by a factor of 2 in order to relate it to the number of $U(N)$ gauge groups $G$.
\end{itemize}
Perfect matchings carry charges under the mesonic and baryonic symmetries. The choices of assigning charges on perfect matchings are under certain basic constraints which are summarized in appendix \sref{appch}. 
\\

\noindent\textbf{Mesonic Moduli Space $\mathcal{M}^{mes}$.} The mesonic moduli space is a subspace of the master space. It has the following properties:
\begin{itemize}
\item The \textbf{mesonic moduli space} for the one D3 brane theory is the following symplectic quotient
\beal{es2c_20}
\mathcal{M}^{mes}=(\mathbb{C}^c[p_1,\dots,p_c]//Q_F)//Q_D = \mathcal{F}^\flat // Q_D~~.
\eea
\item The mesonic moduli space is a toric Calabi-Yau $3$-fold and is generally lower dimensional than the master space.
\end{itemize}

\noindent\textbf{Hilbert Series.} The Hilbert series is a generating function which counts chiral gauge invariant operators. It contains information on moduli space generators and their relations. A method known as \textbf{plethystics} \cite{Benvenuti:2006qr,Feng:2007ur,Hanany:2007zz} is used to extract the information from the Hilbert series. For charges $Q$ which are either $Q_F$ or $Q_t$ for $\master$ and $\mesonic$ respectively, the corresponding Hilbert series is given by the \textbf{Molien Integral}
\beal{es2c_30}
g_1(y_\alpha;\mathcal{M}) = \prod_{i=1}^{|Q|} \oint_{|z_i|=1} \frac{\ud z_i}{2\pi i z_i}
\prod_{\alpha=1}^{c} \frac{1}{1-y_\alpha \prod_{j=1}^{|Q|} z_j^{{Q}_{j\alpha}}}
\eea
where $c$ is the number of perfect matchings and $|Q|$ is the number of rows in the charge matrix $Q$. The fugacity $y_\alpha=t_\alpha$ counts extremal perfect matchings and the fugacity $y_{s_m}$ counts all other fugacities $s_m$.
\\

\section{An introduction to Specular Duality \label{s3}}

The following section reviews toric duality of brane tilings and compares it with specular duality. The section illustrates how the new correspondence is related to the untwisting map \cite{Feng:2005gw,Franco:2011sz,Stienstra:2007dy} and the shiver \cite{Butti:2007jv,Franco:2007ii,Hanany:2008fj}. We focus on the 30 brane tilings with reflexive toric diagrams.
\\

\subsection{Toric Duality and Specular Duality \label{s2d}}

\noindent\textbf{Toric Duality.}
Two $4d$ quiver gauge theories with brane tilings are called toric dual \cite{Feng:2000mi,Feng:2001xr,Feng:2002zw,Seiberg:1994pq,Feng:2001bn,2001JHEP...12..001B,Franco:2003ea} if in the UV they have different Lagrangians with a different field content and superpotential, but flow to the same universality class in the IR.

\begin{figure}[ht!]
\begin{center}
\includegraphics[trim=0cm 0cm 0cm 0cm,width=12 cm]{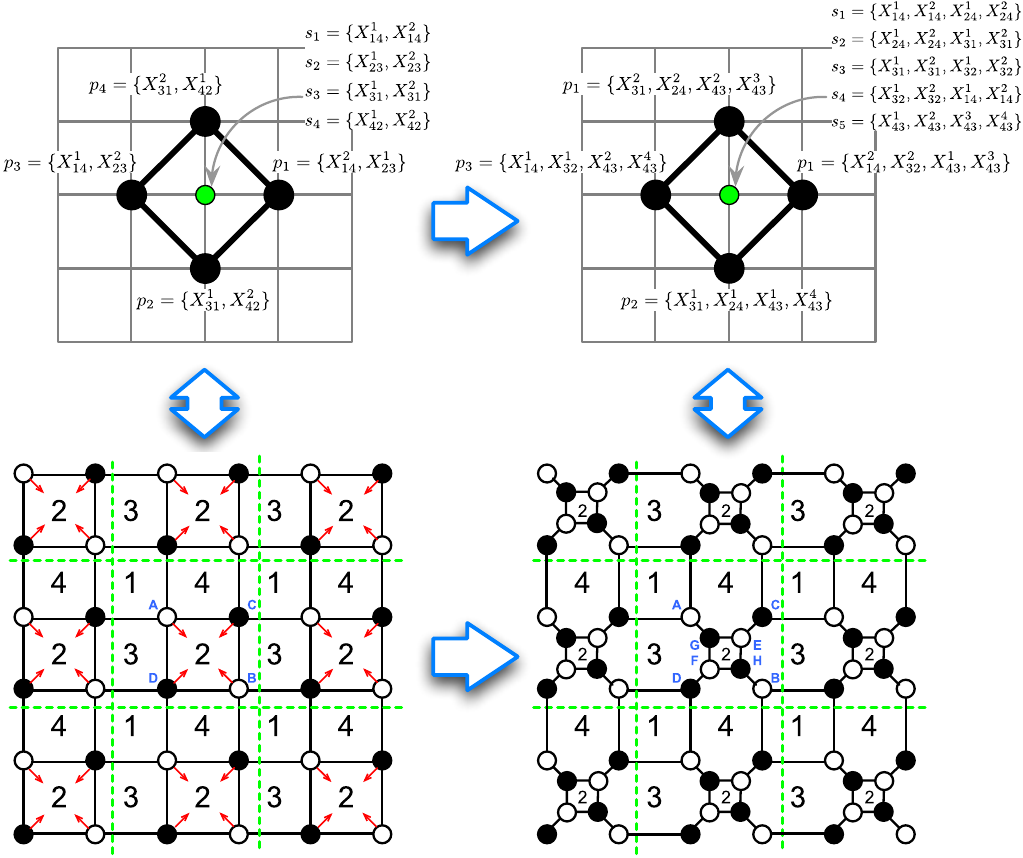}
\caption{\textit{`Urban Renewal'.} Toric duality acts on the brane tiling of the zeroth Hirzebruch surface $F_0$. The points in the toric diagram correspond to perfect matchings and GLSM fields. Perfect matchings are defined as sets of quiver fields.}
  \label{fseiberg}
 \end{center}
 \end{figure}

Let us summarise the properties of toric duality for brane tilings:
\begin{itemize}

\item The \textit{mesonic moduli spaces} $\mathcal{M}^{mes}$ are the same, but the master spaces $\master$ are not \cite{Forcella:2008ng}. The mesonic Hilbert series are the same up to a fugacity map.

\item The \textit{toric diagrams} of $\mathcal{M}^{mes}$ are $GL(2,\mathbb{Z})$ equivalent. However, multiplicities of internal toric points with zero R-charge can differ.

\item The number of \textit{gauge groups} $G$ remains constant.

\end{itemize}
\vspace{0.5cm}

\noindent\textit{Example.} The Hirzebruch $\mathbb{F}_0$ model has the superpotential
\beal{es2c_50}
W_{I} =
  \underbracket[0.1mm]{X_{14}^{1} X_{42}^{1} X_{23}^{1} X_{31}^{1}}_{A}
+ \underbracket[0.1mm]{X_{14}^{2} X_{42}^{2} X_{23}^{2} X_{31}^{2}}_{B}
- \underbracket[0.1mm]{X_{14}^{2} X_{42}^{1} X_{23}^{2} X_{31}^{1}}_{C}
- \underbracket[0.1mm]{X_{14}^{1} X_{42}^{2} X_{23}^{1} X_{31}^{2}}_{D}
~~,
\eea
with the corresponding brane tiling and toric diagram shown in the left panel of \fref{fseiberg}. By dualizing on the gauge group $U(N)_2$, the superpotential becomes
\beal{es2c_51}
W_{II} &=&
  \underbracket[0.1mm]{X_{14}^{1}X_{43}^{1}X_{31}^{1}}_{A}
+ \underbracket[0.1mm]{X_{14}^{2}X_{43}^{2}X_{31}^{2}}_{B}
- \underbracket[0.1mm]{X_{14}^{2}X_{43}^{3}X_{31}^{1}}_{C}
- \underbracket[0.1mm]{X_{14}^{1}X_{43}^{4}X_{31}^{2}}_{D}
\nn\\
&&
+ \underbracket[0.1mm]{X_{14}^{1}X_{43}^{3}X_{31}^{2}}_{E}
+ \underbracket[0.1mm]{X_{14}^{2}X_{43}^{4}X_{31}^{1}}_{F}
- \underbracket[0.1mm]{X_{14}^{1}X_{43}^{1}X_{31}^{1}}_{G}
- \underbracket[0.1mm]{X_{14}^{2}X_{43}^{2}X_{31}^{2}}_{H}
\eea
with the associated brane tiling and toric diagram shown in the right panel of \fref{fseiberg}. The figure labels toric points with the associated perfect matchings.
\\

\noindent\textbf{Specular Duality.}  The new correspondence has the following properties for dual brane tilings:
\begin{itemize}
\item $\master$ are isomorphic\footnote{Note: Specular duality extends to the full master space $\mathcal{F}^\flat$. We restrict the discussion to its largest component $\master$.} and the Hilbert series are the same up to a fugacity map.
\item Except for self-dual cases, $\mesonic$ are not the same.
\item The number of gauge groups $G$ remains invariant.
\item The number of matter fields $E$ remains invariant.
\end{itemize}

\begin{figure}[H]
\begin{center}
\resizebox{0.93\hsize}{!}{
\includegraphics[trim=0cm 0cm 0cm 0cm,totalheight=18 cm]{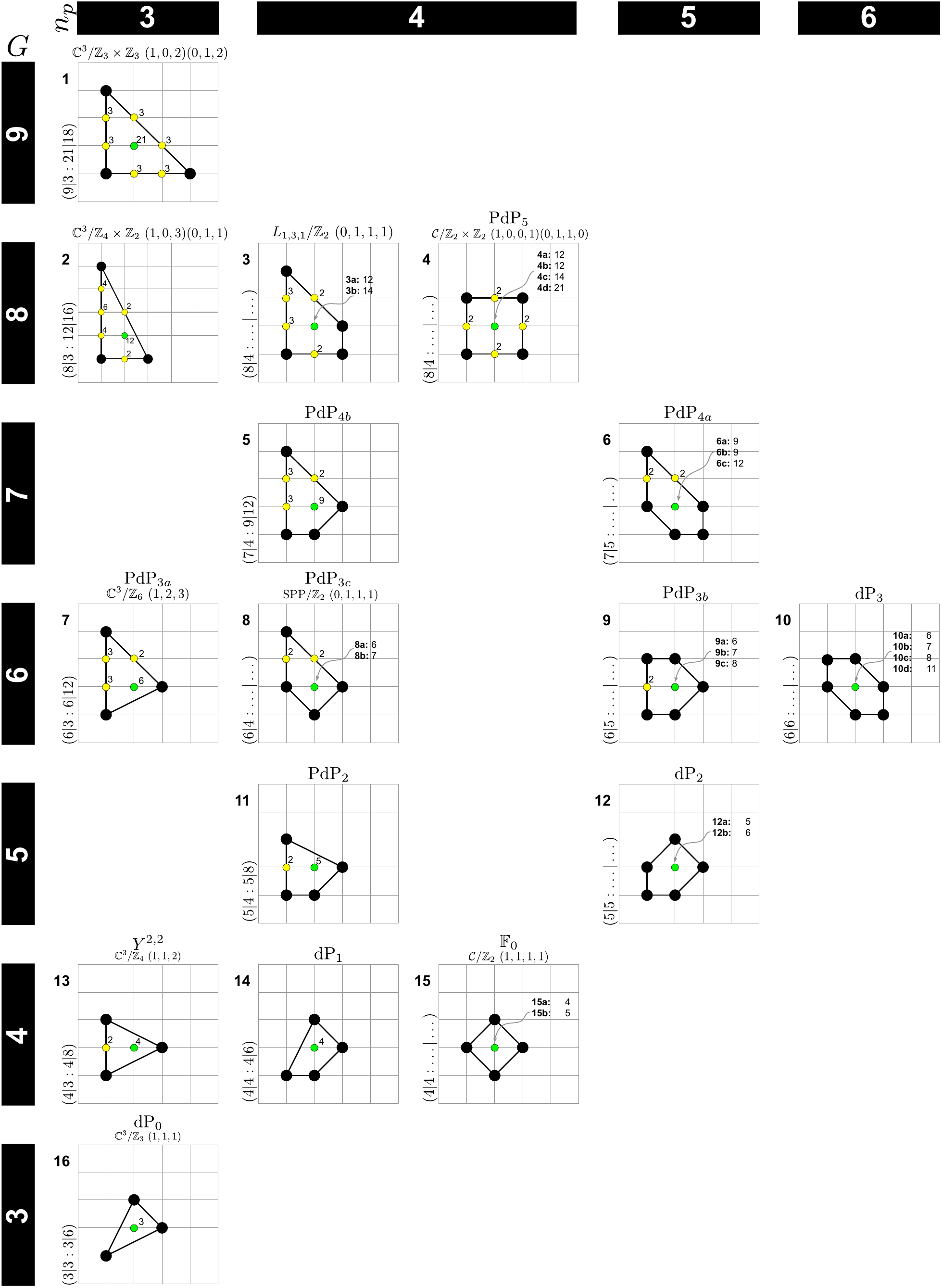}
}
  \caption{\textit{Reflexive Toric Diagrams.} The figure shows the $16$ reflexive toric diagrams which correspond to $30$ brane tilings. Each polygon is labelled by $(G|n_p:n_i|n_W)$, where $G$ is the number of $U(N)$ gauge groups, $n_p$ is the number of extremal perfect matchings, $n_i$ is the number of internal perfect matchings, and $n_W$ is the number of superpotential terms. A reflexive polygon can correspond to multiple brane tilings by toric duality.  \label{f_sumtoric2}}
 \end{center}
 \end{figure}

There are 16 reflexive toric diagrams. They are summarized in \fref{f_sumtoric2} \cite{Hanany:2012hi} and relate to 30 brane tilings. Specular duality exhibits additional properties for this set of brane tilings:
\begin{itemize}
\item Internal/external perfect matchings of brane tiling A become external/internal perfect matchings of the dual brane tiling B.
\item The mesonic flavour symmetries of brane tiling A become the anomalous or enhanced hidden baryonic symmetries of brane tiling B. 
\end{itemize}

\begin{table}[H]
\centering
\begin{tabular}{|c|c|}
\hline
$d$ & Number of Polytopes\\
\hline\hline
1 & 1 \\
\hline
2 & 16\\
\hline 
3 & 4319\\
\hline
4 & 473800776\\
\hline
\end{tabular}
\caption{\textit{Counting Reflexive Polytopes.} Number of distinct reflexive lattice polytopes in dimension $d\leq 4$. The number of polytopes forms a sequence which has the OEIS identifier A090045.} \label{tpolycount}
\end{table}

\noindent
As noted above, specular duality exhibits additional properties for brane tilings with reflexive toric diagrams. Many of the 30 brane tilings which correspond to the 16 reflexive polygons are toric duals \cite{Hanany:2012hi}.

Reflexive polytopes have the following properties:
\begin{itemize}
\item A \textbf{reflexive polytope} is a convex $\mathbb{Z}^{d}$ lattice polytope whose unique interior point is the origin $(0,\dots,0)$.

\item A \textbf{dual (polar) polytope} exists for every reflexive polytope $\Delta$. The dual $\Delta^{\circ}$ is another lattice polytope with points
\beal{es00_20}
\Delta^{\circ}=\{
v^{\circ}\in\mathbb{Z}^d ~|~ \langle v^{\circ},v \rangle \geq -1 ~\forall v\in \Delta
\}
\eea
$\Delta^{\circ}$ is another reflexive polytope. There are self-dual polytopes, $\Delta=\Delta^{\circ}$.\footnote{Note that this duality between reflexive polytopes does not correspond to specular duality.}

\item A \textbf{classification of reflexive polytopes} \cite{Kreuzer:1998vb,Kreuzer:2000qv,Kreuzer:2000xy} is available for the dimensions $d\leq 4$ as shown in \tref{tpolycount}. 
\end{itemize}
\vspace{0.5cm}

\begin{figure}[ht!]
\begin{center}
\resizebox{1\hsize}{!}{
\includegraphics[trim=0cm 0cm 0cm 0cm,totalheight=18 cm]{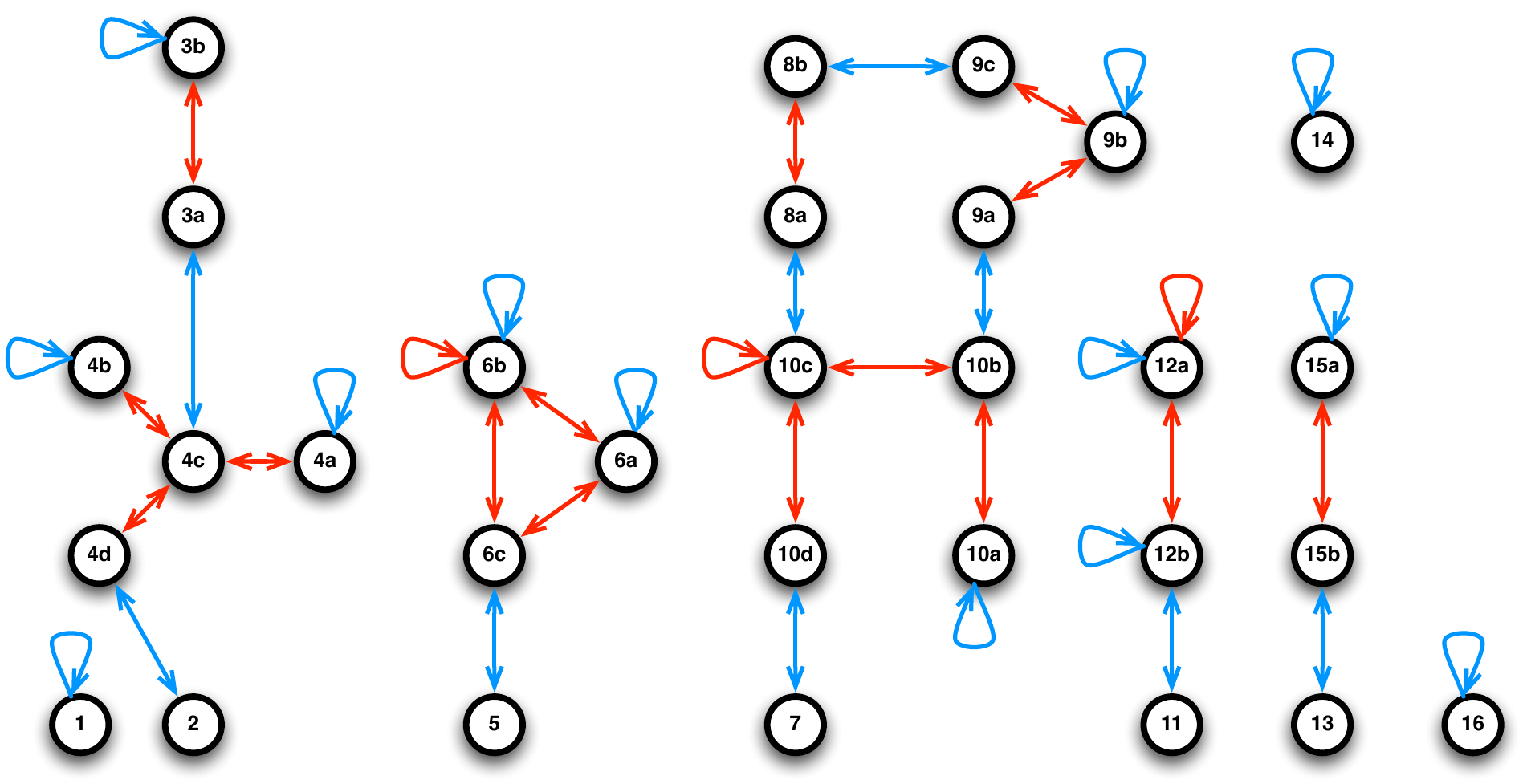}
}
  \caption{\textit{Toric and Specular Duality.} These are the duality trees of brane tilings (nodes) with reflexive toric diagrams. The brane tiling labels are taken from \cite{Hanany:2012hi} and \fref{f_sumtoric2}. Arrows indicate toric duality (red) and specular duality (blue).
  \label{fctree2}}
 \end{center}
 \end{figure}

\begin{landscape}
\begin{figure}[ht!!]
\begin{center}
\includegraphics[trim=0.5cm 0.5cm 0cm 0.5cm,totalheight=14.6 cm]{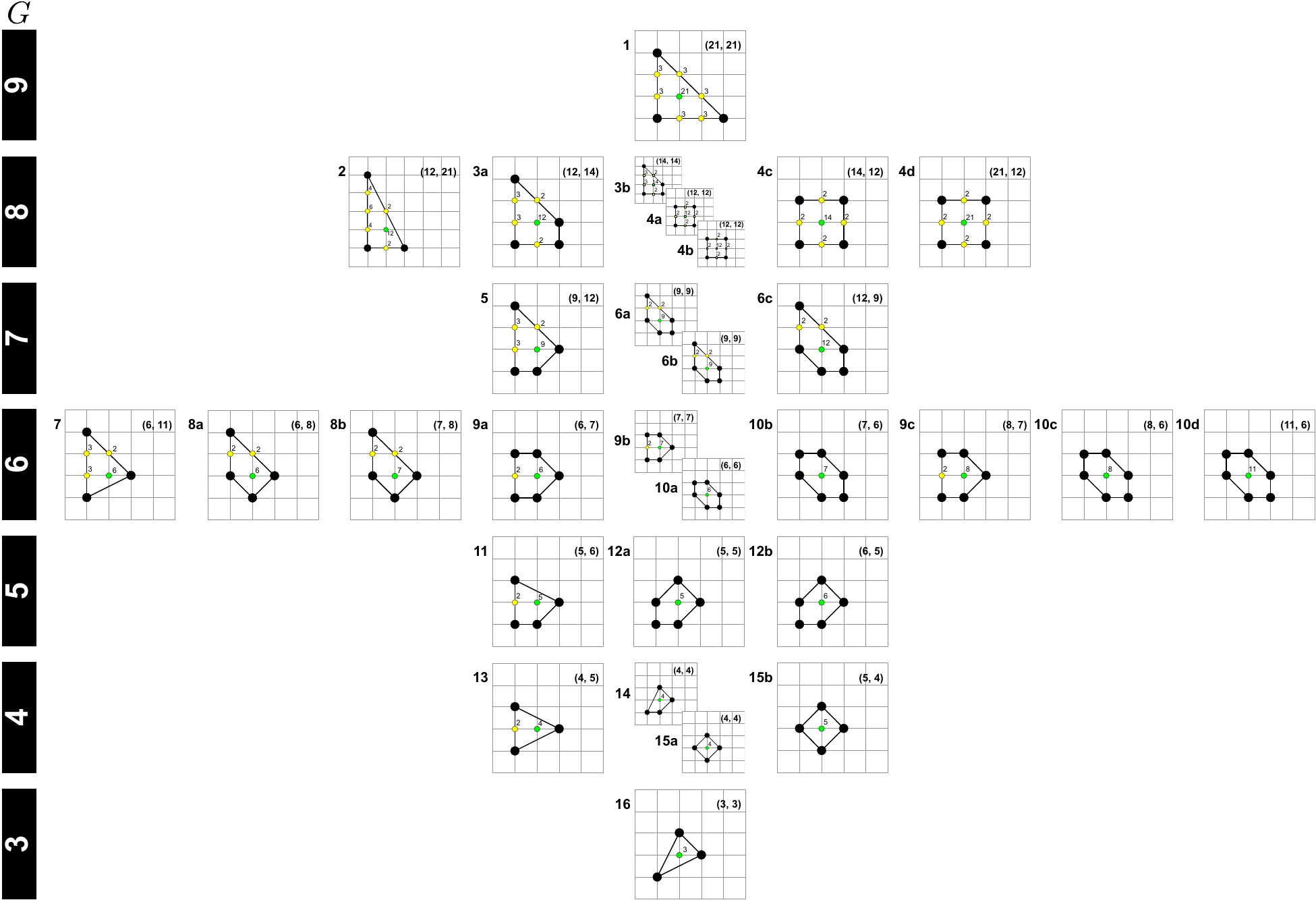}
\end{center}
\caption{\textit{Arbor specularis}. The 30 reflexive toric diagrams with perfect matching multiplicities. The models are labelled with $(n_i,n_e)$, where $n_i$ and $n_e$ are the number of internal and external perfect matchings respectively. The $y$-axis is labelled by the number of gauge groups $G$ or the area of the polygon, and the position along the $x$-axis relates to the difference $n_i-n_e$. \label{fspecdual}}
\end{figure}
\end{landscape}

Specular duality preserves the reflexivity of the toric diagram and the set of $30$ brane tilings in \fref{f_sumtoric2}:
\beal{es2c_40}
&
1 \leftrightarrow 1
&
\nn\\
&
2 \leftrightarrow 4d
~,~
3a \leftrightarrow 4c
~,~
3b \leftrightarrow 3b
~,~
4a \leftrightarrow 4a
~,~
4b \leftrightarrow 4b
&
\nn\\
&
5 \leftrightarrow 6c
~,~
6a \leftrightarrow 6a
~,~
6b \leftrightarrow 6b
&
\nn\\
&
7 \leftrightarrow 10d
~,~
8a \leftrightarrow 10c
~,~
8b \leftrightarrow 9c
~,~
9a \leftrightarrow 10b
~,~
9b \leftrightarrow 9b
~,~
10a \leftrightarrow 10a
&
\nn\\
&
11 \leftrightarrow 12b
~,~
12a \leftrightarrow 12a
&
\nn\\
&
13 \leftrightarrow 15b
~,~
14 \leftrightarrow 14
~,~
15a \leftrightarrow 15a
&
\nn\\
&
16 \leftrightarrow 16
&
~.
\eea
\fref{fspecdual} shows the reflexive toric diagrams with specular dual brane tilings. 
\\

\begin{figure}[ht!!]
\begin{center}
\includegraphics[trim=0.5cm 0.5cm 0cm 0.5cm,totalheight=14.5 cm]{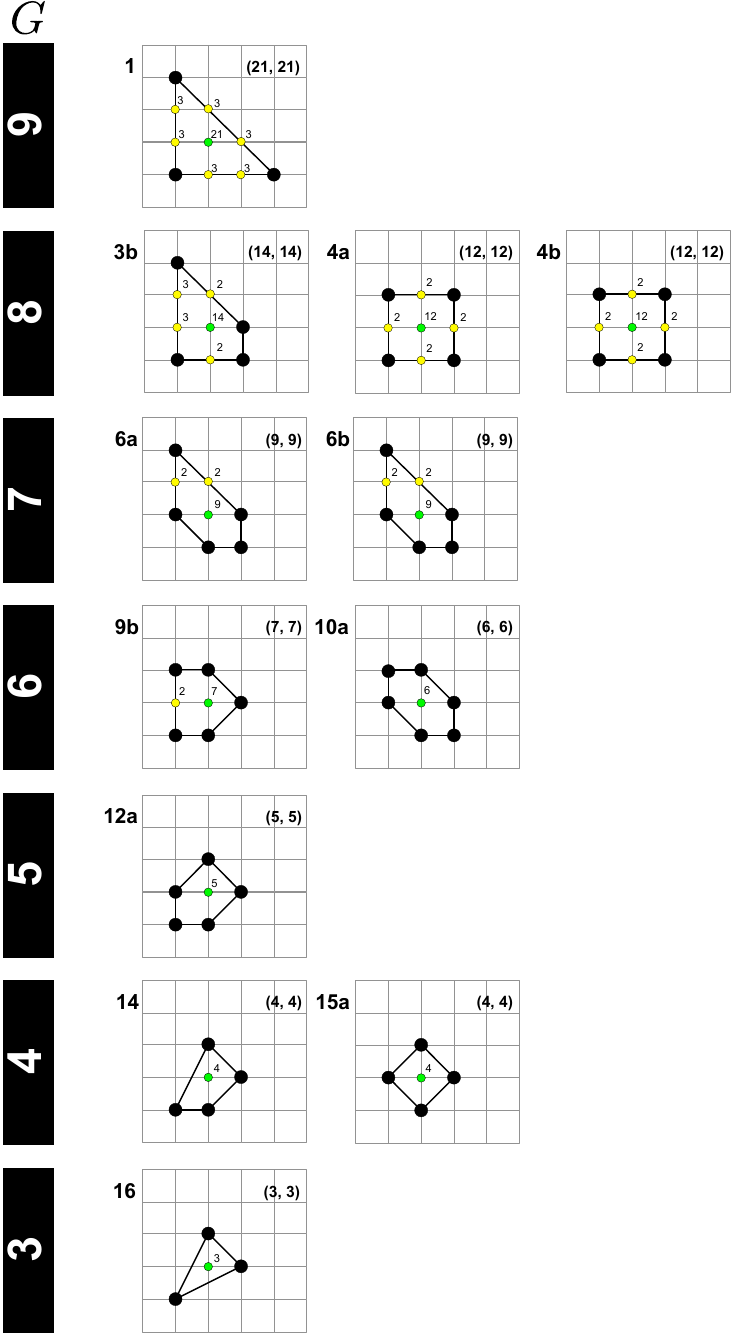}
\end{center}
\caption{\textit{Self-duals under Specular Duality.} These are the 12 reflexive toric diagrams which have self-dual brane tilings. The models are labelled with $(n_i,n_e)$, where $n_i$ and $n_e$ are the number of internal and external perfect matchings respectively. \label{fspecselfdual}}
\end{figure}

\noindent\textbf{Self-dual Brane Tilings.} Certain brane tilings with reflexive toric diagrams are self-dual. These are:
\beal{es2c_41}
1~,~
3b~,~ 
4a~,~ 
4b~,~
6a~,~
6b~,~
9b~,~
10a~,~
12a~,~
14~,~
15a~,~
16~,
\eea
which are summarized in \fref{fspecselfdual}.
The toric diagram and brane tiling are invariant under specular duality.
\\

\subsection{Specular Duality and `Fixing' Shivers \label{s4}}
 
As illustrated in Section \sref{s2d}, toric duality has a natural interpretation on the brane tiling. The following section identifies the interpretation of specular duality on the brane tiling.

A toric singularity has an associated \textbf{characteristic polynomial}, also known as the Newton polynomial,
\beal{es4_1}
P(w,z)= \sum_{(n_1,n_2)\in \Delta} a_{n_1,n_2} w^{n_1} z^{n_2} ~~,
\eea
where the sum runs over all points in the toric diagram, and $a_{n_1,n_2}$ is a complex number. The geometric description of the \textbf{mirror manifold} \cite{Hori:2000kt,Hori:2000ck,Feng:2005gw} is
\beal{es4_2}
Y&=&P(w,z)~,~ \nn\\
Y&=& u v~,
\eea
where $w,z\in\mathbb{C}^{*}$ and $u,v\in\mathbb{C}$. The curve $P(w,z)-Y=0$ describes a punctured \textbf{Riemann surface} $\Sigma_Y$ with
\begin{itemize}
\item the genus $g$ corresponding to the number $I$ of internal toric points 
\item the number of punctures corresponding to the number $E$ of external toric points.
\end{itemize}
The Riemann surface is fibered over each point in $Y$. Of particular interest to us is the Riemann surface $\Sigma$ fibered over the origin $Y=0$. It is related to the brane tiling on $\mathbb{T}^2$ under the \textbf{untwisting map} $\phi_u$ \cite{Feng:2005gw,Franco:2011sz,Stienstra:2007dy}.

A brane tiling consists of \textbf{zig-zag paths} $\eta_i$ \cite{2003math.ph...5057K,Hanany:2005ss}. These are collections of edges in the tiling that form closed non-trivial paths on $\mathbb{T}^2$. Zig-zag paths maximally turn left at a black node and then maximally turn right at the next adjacent white node. The \textbf{winding numbers} $(p,q)$ of zig-zag paths relate to the $\mathbb{Z}^2$ direction of the corresponding leg in the $(p,q)$-web \cite{Aharony:1997bh}. The dual of the $(p,q)$-web is the toric diagram. By thickening the $(p,q)$-web, one obtains the punctured Riemann surface $\Sigma$.
 
\begin{figure}[H]
\begin{center}
\begin{tabular}{ccc}
 brane tiling on $\mathbb{T}^2$ & $\stackrel{\phi_u}{\rightarrow}$ & shiver on $\Sigma$\\
 zig-zag path $\eta_i$ & $\mapsto$ & puncture $\gamma_i$\\
 face/gauge group $U(N)_a$ & $\mapsto$ &  zig-zag path $\tilde{\eta}_a$ \\
 node/term $w_k,~b_k$ & $\mapsto$ & node/term $w_k,~b_k$\\
 edge/field $X_{ab}$ & $\mapsto$ & edge/field $X_{ij}$\\
&&\\
\includegraphics[trim=0cm 0cm 0cm 0cm,width=5 cm]{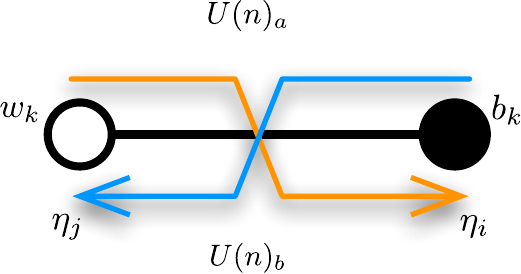} &
&
\includegraphics[trim=0cm 0cm 0cm 0cm,width=5 cm]{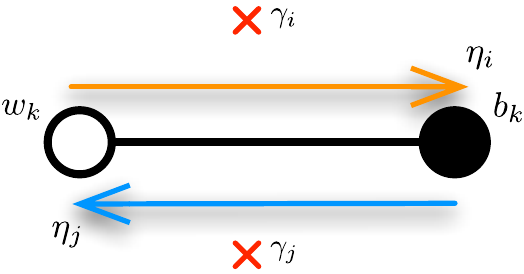}
\end{tabular}
\end{center}
\caption{\textit{The Untwisting Map $\phi_u$.} The untwisting map relates a brane tiling on $\mathbb{T}^2$ to a shiver on a punctured Riemann surface $\Sigma$.\label{funtwist}}
\end{figure}

The untwisting map $\phi_u$ has the following action on the brane tiling:
\beal{esu1}
\phi_u ~:~\hspace{2cm} \text{brane tiling on}~\mathbb{T}^2 &\rightarrow& \text{shiver on}~\Sigma 
\nn\\
\text{zig-zag path}~\eta_i &\mapsto& \text{puncture}~\gamma_i
\nn\\
\text{face/gauge group}~U(N)_a &\mapsto& \text{zig-zag path}~\tilde{\eta}_a
\nn\\
\text{node/term}~w_k,~b_k &\mapsto& \text{node/term}~w_k,~b_k
\nn\\
\text{edge/field}~X_{ab} &\mapsto& \text{edge/field}~X_{ij}~~,
\eea
where $a,b$ count $U(N)$ gauge groups/brane tiling faces, $i,j$ count zig-zag paths on the original brane tiling on $\mathbb{T}^2$, and $\tilde{\eta}_a$ are zig-zag paths of the shiver on $\Sigma$. An illustration of the untwisting map is in \fref{funtwist}.

The untwisted brane tiling on $\Sigma$ is known as a \textbf{shiver} \cite{Butti:2007jv,Franco:2007ii,Hanany:2008fj}. It is not associated to a quiver, a superpotential and a field theory moduli space, and therefore can be interpreted as a `pseudo-brane tiling' on a punctured Riemann surface. An interesting question to ask at this point is whether a shiver can be `fixed' by a map $\phi_f$ such that it becomes a consistent brane tiling. 

The main obstructions are the punctures of $\Sigma$ which have no interpretation in the quiver gauge theory context. Let the punctures therefore be identified with $U(N)$ gauge groups under the following definition of the \textbf{shiver fixing map}:
\beal{esu2}
\phi_f ~:~ \hspace{2cm}
\text{shiver on}~\Sigma &\rightarrow& \text{brane tiling on}~\Sigma
\nn\\
\text{puncture}~\gamma_i &\mapsto& \text{face/gauge group}~U(N)_i~~,
\eea
with the zig-zag paths $\tilde{\eta}_a$, nodes $w_k$ and $b_k$, and edges $X_{ij}$ on the shiver remaining invariant. 

\begin{figure}[H]
\begin{center}
\resizebox{\hsize}{!}{
\begin{tabular}{ccccc}
brane tiling A on $\mathbb{T}^2$ & $\stackrel{\phi_u}{\rightarrow}$ & shiver on $\Sigma$
& $\stackrel{\phi_f}{\rightarrow}$ &  brane tiling B on $\Sigma$
\\
zig-zag path $\eta_i$ & $\mapsto$ & puncture $\gamma_i$
& $\mapsto$ & face/gauge group $U(N)_i$
\\
face/gauge group $U(N)_a$ & $\mapsto$ &  zig-zag path $\tilde{\eta}_a$
& $\mapsto$ & zig-zag path $\tilde{\eta}_a$
\\
node/term $w_k,~b_k$ & $\mapsto$ & node/term $w_k,~b_k$
& $\mapsto$ & node/term $w_k,~b_k$
\\
edge/field $X_{ab}$ & $\mapsto$ & edge/field $X_{ij}$
& $\mapsto$ & edge/field $X_{ij}$
\\
&&\\
\includegraphics[trim=0cm 0cm 0cm 0cm,width=7 cm]{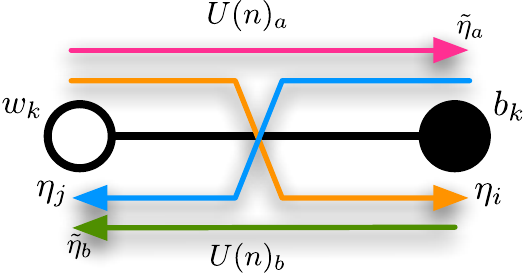} &
&
\includegraphics[trim=0cm 0cm 0cm 0cm,width=7 cm]{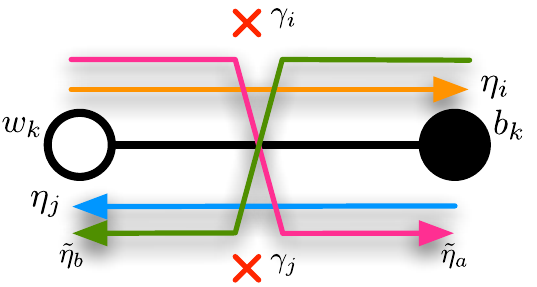}
&&
\includegraphics[trim=0cm 0cm 0cm 0cm,width=7 cm]{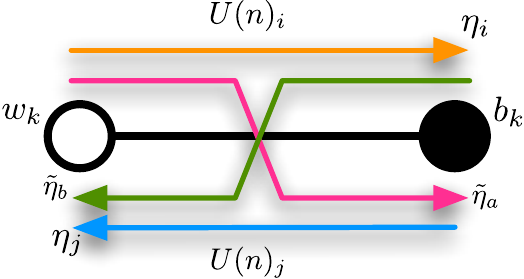}
\end{tabular}
}
\end{center}
\caption{\textit{Specular Duality on a Brane Tiling.} The map $\phi_{\text{specular}}=\phi_f \circ \phi_u$ which defines specular duality first untwists a brane tiling and then replaces punctures with $U(N)$ gauge groups.\label{fspecularmap}}
\end{figure}

Accordingly, using the shiver fixing map $\phi_f$ and the untwisting map $\phi_u$,  \textbf{specular duality} on brane tilings can be defined as follows
\beal{esu3}
\phi_{\text{specular}} =
\phi_f \circ \phi_u 
~:~
\hspace{2cm} \text{brane tiling A on}~\mathbb{T}^2 &\rightarrow& \text{brane tiling B on}~\Sigma
\nn\\
\text{zig-zag path}~\eta_i &\mapsto& \text{face/gauge group}~U(N)_i
\nn\\
\text{face/gauge group}~U(N)_a &\mapsto& \text{zig-zag path}~\tilde{\eta}_a
\nn\\
\text{node/term}~w_k,~b_k &\mapsto& \text{node/term}~w_k,~b_k
\nn\\
\text{edge/field}~X_{ab} &\mapsto& \text{edge/field}~X_{ij}~~,
\eea
where $\phi_{\text{specular}}$ is invertible. A graphical illustration of $\phi_{\text{specular}}$ is in \fref{fspecularmap}.

For a brane tiling to have a Calabi-Yau 3-fold as its mesonic moduli space and to have a known AdS dual \cite{Benvenuti:2004dw,Benvenuti:2005wi,Kennaway:2007tq}, it needs to be on $\mathbb{T}^2$. Brane tilings with reflexive toric diagrams have a specular dual which is always on $\Sigma=\mathbb{T}^2$. This is because, as we recall, reflexive toric diagrams always have by definition $I=1$ and their $(p,q)$-web has therefore always genus $g=1$. 
\\

\begin{figure}[H]
\begin{center}
\includegraphics[trim=0cm 0cm 0cm 0cm,width=15 cm]{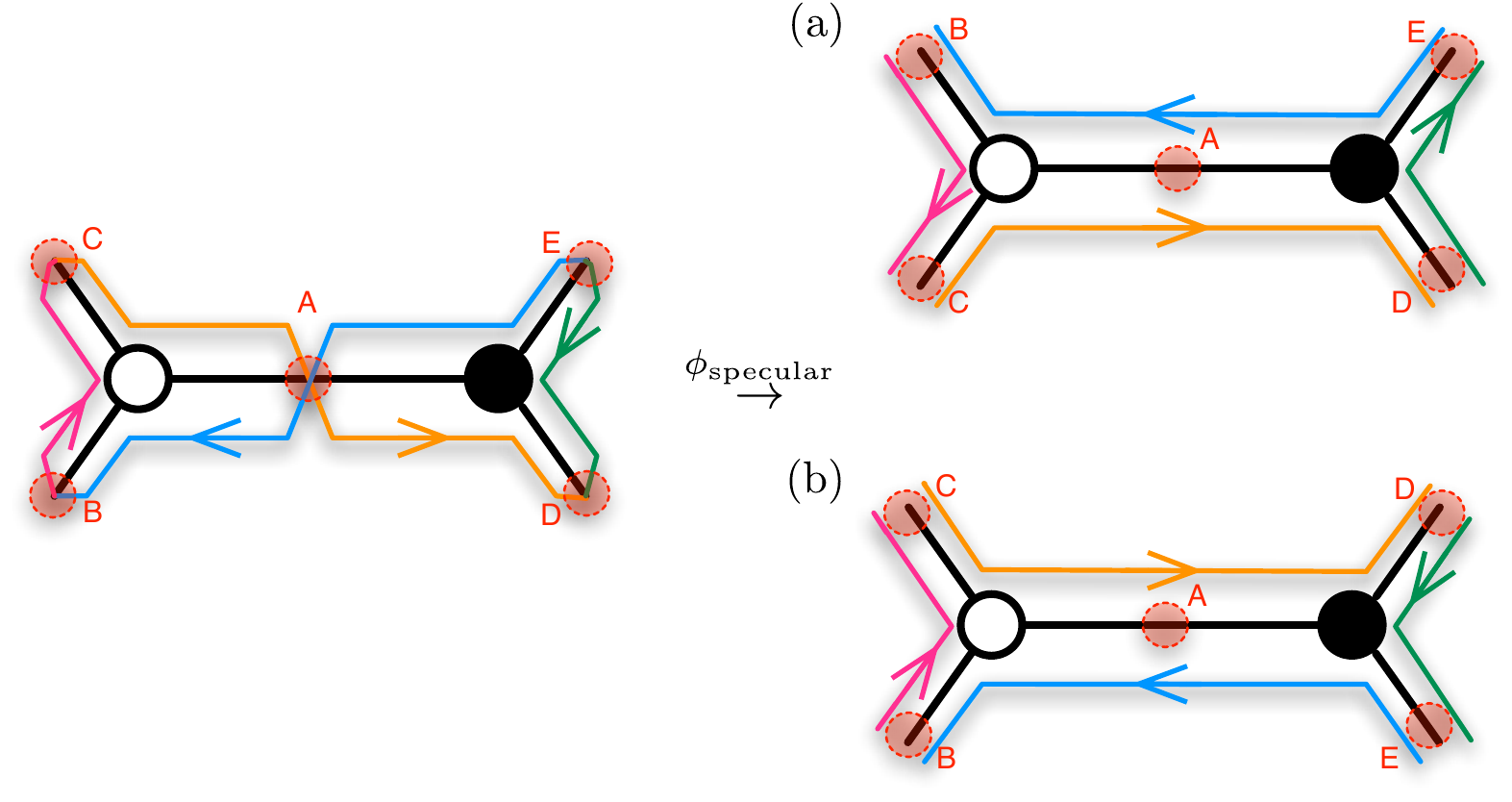}
\end{center}
\caption{\textit{Untwisting the Superpotential.} There are two equivalent ways of untwisting the brane tiling. The order of fields around either a white (clockwise) or black (anti-clockwise) node in the brane tiling is reversed under the untwisting. Either way results in the same brane tiling.
\label{fcyclic}}
\end{figure}

\noindent\textbf{Invariance of the master space $\master$.} Specular duality has an important effect on a brane tiling's superpotential $W$ which can be demonstrated with the following example
\beal{esuu1}
W = \dots + A B C - A D E + \dots ~~,
\eea
where $A,\dots,E$ are quiver fields.\footnote{There is an overall trace in the superpotential which is not written down for simplicity.} The corresponding nodes in the brane tiling are illustrated along with zig-zag paths in the left panel of \fref{fcyclic}.

Specular duality untwists the brane tiling in such a way that the order of quiver fields around either white (clockwise) nodes or black (anti-clockwise) nodes is reversed. For the example in \eref{esuu1}, the superpotential of the dual brane tiling has either the form
\beal{esuu2}
W_{\text{(a)}}=
\dots + A C B - A D E + \dots
\eea
or the form
\beal{esuu3}
W_{\text{(b)}}=
\dots + A B C - A E D + \dots
\eea
as illustrated in the right panel of \fref{fcyclic}. The options of reversing the orientation around white nodes or black nodes are equivalent up to an overall swap of node colours.

For the case of single D3 brane theories with $U(1)$ gauge groups, the fields commute such that
\beal{esuu2}
W=W_{\text{(a)}}=W_{\text{(b)}}~~.
\eea
The $U(1)$ superpotential is invariant under specular duality. Since the master space $\master$ is defined in terms of F-terms, the observation in \eref{esuu2} implies that it is invariant under specular duality.
\\

\noindent\textbf{No specific Quiver from an Abelian $W$.} In order to show that the master spaces of dual one brane theories are isomorphic, it is sufficient to illustrate that the superpotentials are the same when the quiver fields commute. However, it is important to note that if the cyclic order of fields in a given superpotential is not recorded, its correspondence to a specific quiver and hence a brane tiling is not unique. A simple example would be the Abelian potential for $\mathbb{C}^3$ or the conifold $\mathcal{C}$ which is $W=0$. In contrast to the distinct non-Abelian superpotentials, the trivial Abelian superpotential for these models encodes no information about the field content of the associated brane tilings.

Since specular duality is a well defined map between brane tilings, not just between Abelian superpotentials, we study in the following sections the new correspondence with the help of \textit{characteristics of the mesonic moduli space}. An important observation is that specular duality exchanges internal and external perfect matchings for brane tilings with reflexive toric diagrams. The difference between internal and external perfect matchings is a property of the mesonic moduli space and its toric diagram.

Perfect matchings as GLSM fields are used for the symplectic quotient description of $\master$. Since perfect matchings represent a choice of coordinates to identify the master space cone, one is free to introduce a new set of coordinates that correspond to the global symmetry of the field theory. In the following sections, we identify coordinate transformations that relate the exchange of internal and external perfect matchings to the exchange of mesonic flavour symmetries and hidden or anomalous baryonic symmetries. Moreover, one can find a third set of coordinates which relate to the boundaries of the Calabi-Yau cone and are used to illustrate how an exchange of internal and external perfect matchings leads to a reflection of the $\master$ cone along a hyperplane.
\\

\section{Model 13 ($Y^{2,2}$, $\mathbb{F}_{2}$, $\mathbb{C}^3/\mathbb{Z}_4$) and Model 15b ($Y^{2,0}$, $\mathbb{F}_{0}$, $\mathcal{C}/\mathbb{Z}_{2}$) \label{s4}}

In the following section, we study specular duality with Model 13 which is known as $Y^{2,2}$, $\mathbb{F}_2$ or $\mathbb{C}^3/\mathbb{Z}_4$ with action $(1,1,2)$ in the literature, and Model 15b which is known as phase II of $Y^{2,0}$, $\mathbb{F}_0$ or $\mathcal{C}/\mathbb{Z}_2$ with action $(1,1,1,1)$.

\subsection{Brane Tilings and Superpotentials \label{s5_1}}

\begin{figure}[ht!]
\begin{center}
\includegraphics[trim=0cm 0cm 0cm 0cm,width=16 cm]{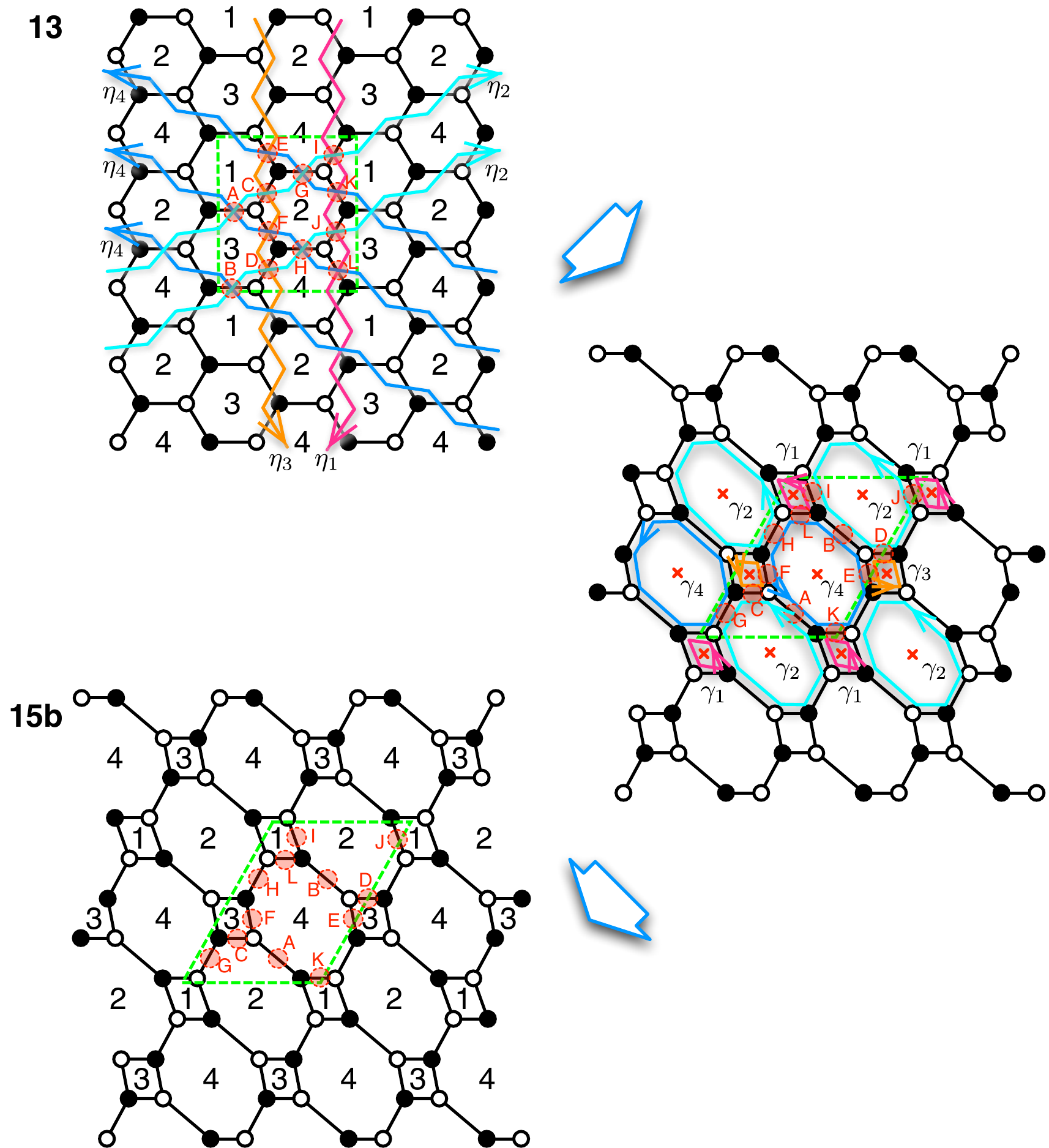}
\end{center}
\caption{
\textit{Specular Duality between Models 13 and 15b.} The untwisting map $\phi_u$ acts on the brane tiling of Model 13 which results in a shiver. The shiver is then fixed with $\phi_f$ which results in the brane tiling of Model 15b.
\label{fm13d}}
\end{figure}

\fref{fm13d} shows how the untwisting map $\phi_u$ acts on the brane tiling of Model 13 to give a shiver. The fixing map $\phi_f$ then takes the shiver to give the brane tiling of Model 15b. Beginning with the superpotential of Model 13,
\beal{esd13_1}
W_{13}&=&
+ X_{12}^{1} X_{24} X_{41}^{1} 
+ X_{31} X_{12}^{2} X_{23}^{2} 
+ X_{41}^{2} X_{13} X_{34}^{1}
+ X_{34}^{2} X_{42} X_{23}^{1} 
\nn\\
&&
- X_{12}^{1} X_{23}^{1} X_{31} 
- X_{13} X_{34}^{2} X_{41}^{1} 
- X_{41}^{2} X_{12}^{2} X_{24} 
- X_{34}^{1} X_{42} X_{23}^{2} 
~~,
\eea
the zig-zag paths are identified as follows
\beal{esd13_2}
\eta_1 &=& \{X_{12}^{1}, X_{23}^{1}, X_{34}^{2}, X_{41}^{1}\}~~,
\nn\\
\eta_2 &=& \{X_{12}^{2}, X_{24}, X_{41}^{1}, X_{13}, X_{34}^{1}, X_{42}, X_{23}^{1}, X_{31}\}~~,
\nn\\
\eta_3 &=& \{X_{23}^{2}, X_{34}^{1}, X_{41}^{2}, X_{12}^{2}\}~~,
\nn\\
\eta_4 &=& \{X_{13}, X_{34}^{2}, X_{42}, X_{23}^{2}, X_{31}, X_{12}^{1}, X_{24}, X_{41}^{2}\}
~~.
\eea
The intersections of zig-zag paths highlighted in \fref{fm13d} are 
\beal{esd13_2}
&&
(A,B,C,D,E,F,G,H,I,J,K,L)=
\nn\\
&&
\hspace{2cm}
(
X_{31},X_{13},X_{12}^{2},X_{34}^{1},X_{41}^{2},X_{23}^{2},X_{24},X_{42},
X_{41}^{1},X_{23}^{1},X_{12}^{1},X_{34}^{2}
)~~.
\eea
Under specular duality, the intersections are mapped to the ones for zig-zag paths on the brane tiling of Model 15b.

In terms of intersections, the superpotential in \eref{esd13_1} takes the form
\beal{esd13_3}
W_{13} &=&
+ K G I + A C F + E B D + L H J \nn\\
&&
- K J A - B L I - E C G - D H F
\eea
The intersections are also fields in the dual brane tiling of Model 15b. Accordingly, the corresponding superpotential can be written as
\beal{esd13_4}
\widetilde{W_{13}} = W_{15b} &=&
+ X_{14}^{1} X_{42}^{1} X_{21}^{1} 
+ X_{42}^{4} X_{23}^{2} X_{34}^{1} 
+ X_{34}^{2} X_{42}^{3} X_{23}^{1}
+ X_{14}^{2} X_{42}^{2} X_{21}^{2} 
\nn\\
&&
- X_{14}^{1} X_{42}^{4} X_{21}^{2} 
- X_{42}^{3} X_{21}^{1} X_{14}^{2} 
- X_{34}^{2} X_{42}^{1} X_{23}^{2} 
- X_{23}^{1} X_{34}^{1} X_{42}^{2} 
\nn\\
&=&
+ K G I 
+ A C F 
+ E B D  
+ L H J 
\nn\\
&&
- K A J 
- B I L 
- E G C 
- D F H 
~~.
\eea
We note that the two superpotentials are the same up to a reversal of cyclic order of negative terms in \eref{esd13_4}. For the Abelian single D3 brane theory, the superpotentials and the corresponding F-terms are the same and hence lead to the same master space $\master$.
\\

\subsection{Perfect Matchings and the Hilbert Series \label{s5_2}}

\begin{figure}[H]
\begin{center}
\includegraphics[trim=0cm 0cm 0cm 0cm,width=4.5 cm]{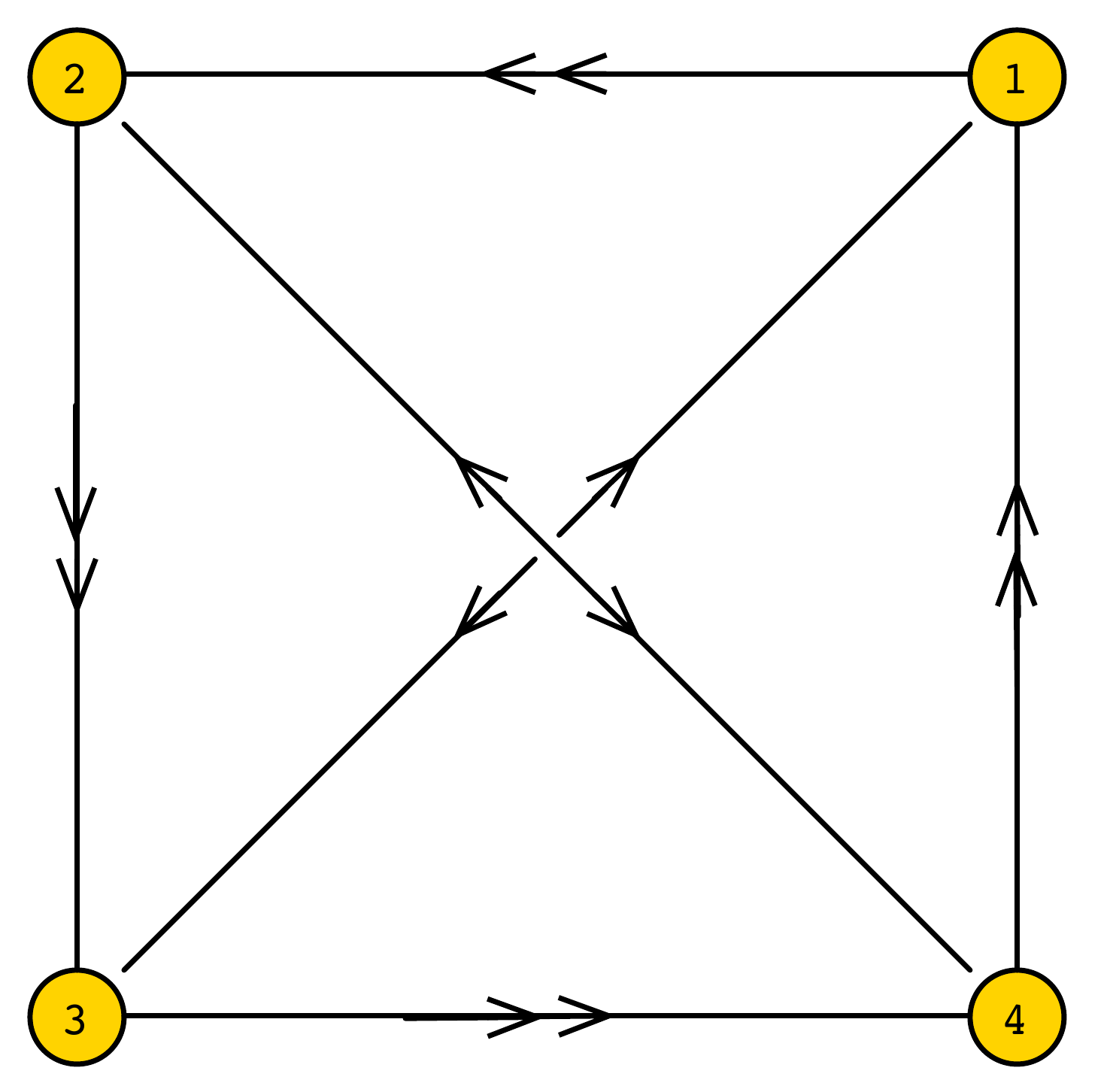}
\includegraphics[width=5 cm]{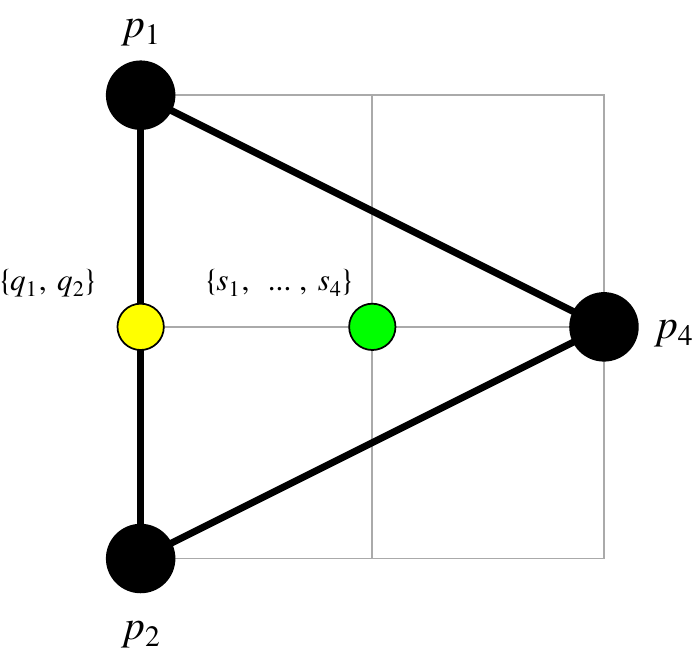}
\includegraphics[width=5 cm]{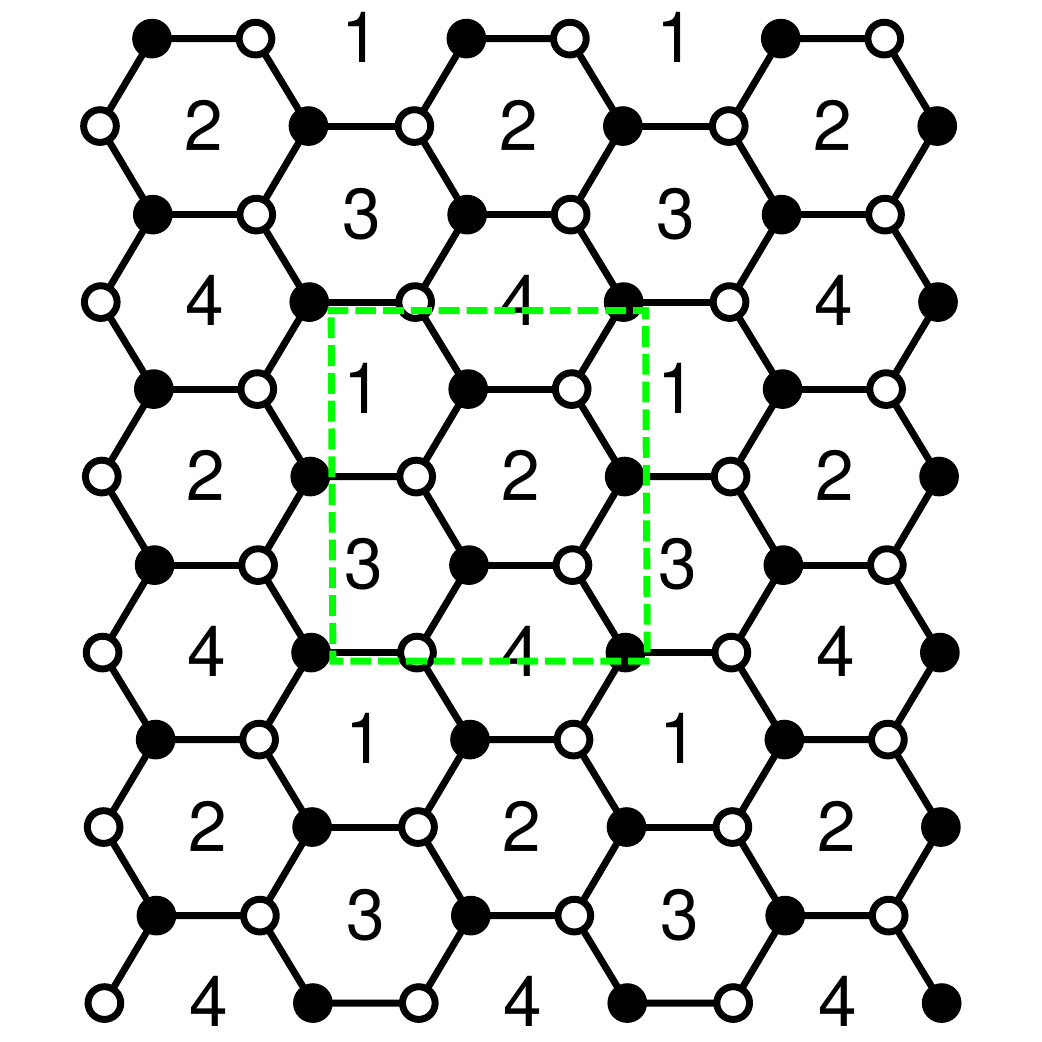}
\\
\vspace{-0.5cm}
\bea
W_{13}&=&
+ X_{12}^{1} X_{24} X_{41}^{1} 
+ X_{31} X_{12}^{2} X_{23}^{2} 
+ X_{41}^{2} X_{13} X_{34}^{1}
+ X_{34}^{2} X_{42} X_{23}^{1} 
\nn\\
&&
- X_{12}^{1} X_{23}^{1} X_{31} 
- X_{13} X_{34}^{2} X_{41}^{1} 
- X_{41}^{2} X_{12}^{2} X_{24} 
- X_{34}^{1} X_{42} X_{23}^{2} 
\nn
\eea
\vspace{-1cm}
\caption{The quiver, toric diagram, brane tiling and superpotential of Model 13.}
  \label{f13}
 \end{center}
 \end{figure}
 
\begin{figure}[H]
\begin{center}
\includegraphics[trim=0cm 0cm 0cm 0cm,width=4.5 cm]{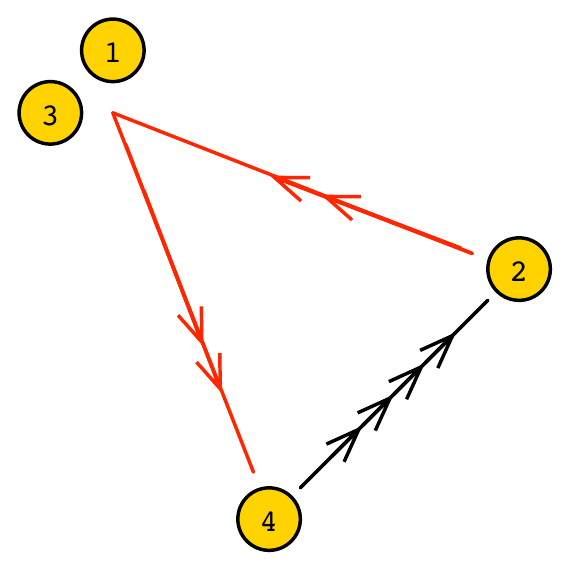}
\includegraphics[width=5 cm]{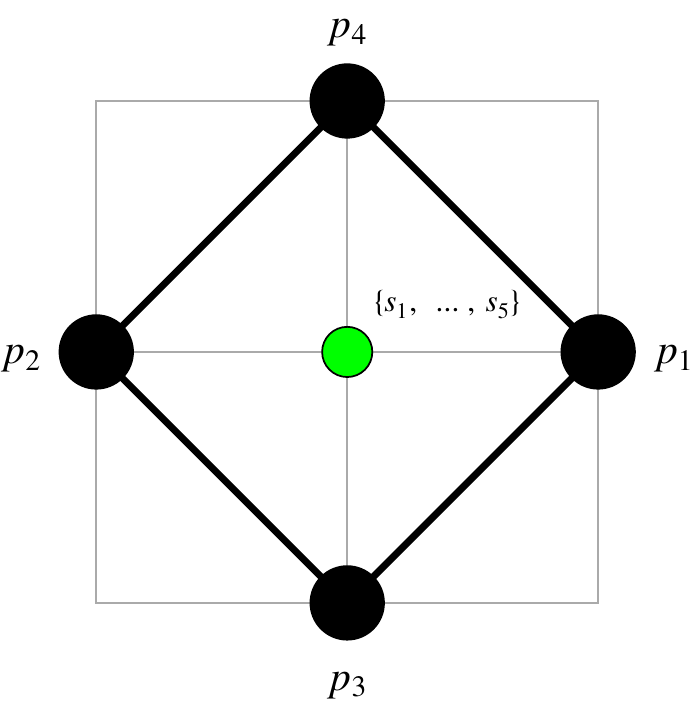}
\includegraphics[width=5 cm]{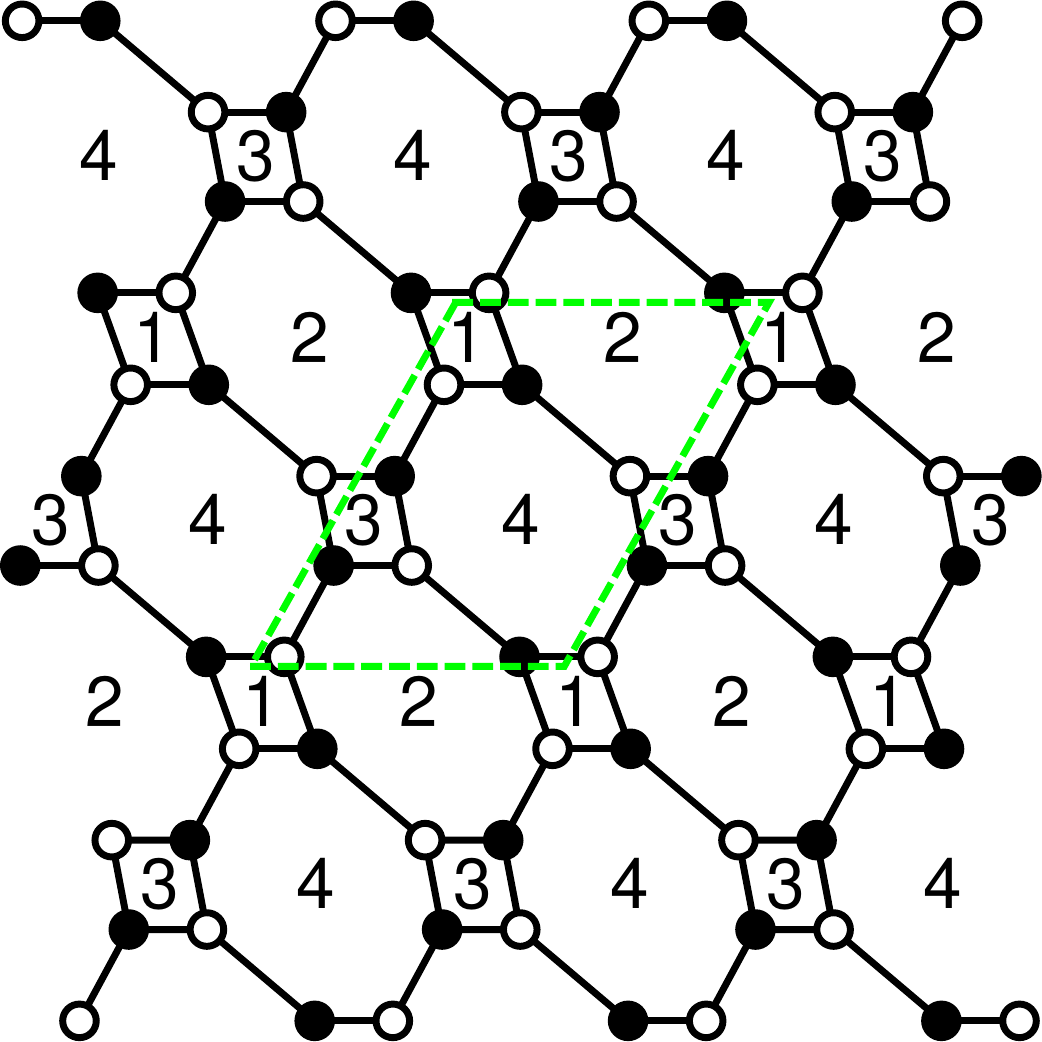}
\\
\vspace{-0.5cm}
\bea
W_{15b} &=&
+ X_{14}^{1} X_{42}^{1} X_{21}^{1} 
+ X_{42}^{4} X_{23}^{2} X_{34}^{1} 
+ X_{34}^{2} X_{42}^{3} X_{23}^{1}
+ X_{14}^{2} X_{42}^{2} X_{21}^{2} 
\nn\\
&&
- X_{14}^{1} X_{42}^{4} X_{21}^{2} 
- X_{42}^{3} X_{21}^{1} X_{14}^{2} 
- X_{34}^{2} X_{42}^{1} X_{23}^{2} 
- X_{23}^{1} X_{34}^{1} X_{42}^{2} 
\nn
\eea
\vspace{-1cm}
\caption{The quiver, toric diagram, brane tiling and superpotential of Model 15b.}
  \label{f15b}
 \end{center}
 \end{figure}

In order to illustrate that specular duality exchanges internal and external perfect matchings of brane tilings, we consider the symplectic quotient description of $\master$. It uses GLSM fields which relate to perfect matchings in a brane tiling. They are summarized in matrices which are for Model 13 and 15b respectively 
 
\noindent\makebox[\textwidth]{%
\footnotesize
$P^{13}=
\left(
\begin{array}{c|ccc|cc|cccc}
 \; & p_{1} & p_{2} & p_{3} & q_{1} & q_{2} & s_{1} &
   s_{2} & s_{3} & s_{4} \\
   \hline
 I=X_{41}^{1} & 1 & 0 & 0 & 1 & 0 & 1 & 0 & 0 & 0 \\
 E=X_{41}^{2} & 0 & 1 & 0 & 1 & 0 & 1 & 0 & 0 & 0 \\
 J=X_{23}^{1} & 1 & 0 & 0 & 1 & 0 & 0 & 1 & 0 & 0 \\
 F=X_{23}^{2} & 0 & 1 & 0 & 1 & 0 & 0 & 1 & 0 & 0 \\
 C=X_{12}^{2} & 1 & 0 & 0 & 0 & 1 & 0 & 0 & 1 & 0 \\
 K=X_{12}^{1} & 0 & 1 & 0 & 0 & 1 & 0 & 0 & 1 & 0 \\
 D=X_{34}^{1} & 1 & 0 & 0 & 0 & 1 & 0 & 0 & 0 & 1 \\
 L=X_{34}^{2} & 0 & 1 & 0 & 0 & 1 & 0 & 0 & 0 & 1 \\
 H=X_{42} & 0 & 0 & 1 & 0 & 0 & 1 & 0 & 1 & 0 \\
 A=X_{31} & 0 & 0 & 1 & 0 & 0 & 1 & 0 & 0 & 1 \\
 B=X_{13} & 0 & 0 & 1 & 0 & 0 & 0 & 1 & 1 & 0 \\
 G=X_{24} & 0 & 0 & 1 & 0 & 0 & 0 & 1 & 0 & 1 
\end{array}
\right)~,~
P^{15b}=
\left(
\begin{array}{c|cccc|ccccc}
 \; & p_1 & p_2 & p_3 & p_4 & s_1 & s_2 & s_3 & s_4 & s_5 \\
 \hline
 I=X_{21}^{1} & 1 & 0 & 0 & 0 & 1 & 0 & 0 & 1 & 0 \\
 E=X_{34}^{2} & 1 & 0 & 0 & 0 & 0 & 1 & 0 & 1 & 0 \\
 J=X_{21}^{2} & 0 & 1 & 0 & 0 & 1 & 0 & 0 & 1 & 0 \\
 F=X_{34}^{1} & 0 & 1 & 0 & 0 & 0 & 1 & 0 & 1 & 0 \\
 C=X_{23}^{2} & 0 & 0 & 1 & 0 & 1 & 0 & 0 & 0 & 1 \\
 K=X_{14}^{1} & 0 & 0 & 1 & 0 & 0 & 1 & 0 & 0 & 1 \\
 D=X_{23}^{1} & 0 & 0 & 0 & 1 & 1 & 0 & 0 & 0 & 1 \\
 L=X_{14}^{2} & 0 & 0 & 0 & 1 & 0 & 1 & 0 & 0 & 1 \\
 H=X_{42}^{2} & 1 & 0 & 1 & 0 & 0 & 0 & 1 & 0 & 0 \\
 A=X_{42}^{4} & 1 & 0 & 0 & 1 & 0 & 0 & 1 & 0 & 0 \\
 B=X_{42}^{3} & 0 & 1 & 1 & 0 & 0 & 0 & 1 & 0 & 0 \\
 G=X_{42}^{1} & 0 & 1 & 0 & 1 & 0 & 0 & 1 & 0 & 0 
\end{array}
\right)~.
$
}
\\

\vspace{0.2cm}
\noindent
The corresponding F-term charge matrices are

\noindent\makebox[\textwidth]{%
\footnotesize
$
Q_{F}^{13}=
\left(
\begin{array}{ccc|cc|cccc}
 p_{1} & p_{2} & p_{3} & q_{1} & q_{2} & s_{1} & s_{2} & s_{3} & s_{4} \\
   \hline
 0 & 0 & -1 & -1 & 0 & 1 & 1 & 0 & 0 \\
 0 & 0 & -1 & 0 & -1 & 0 & 0 & 1 & 1 \\
 1 & 1 & 0 & -1 & -1 & 0 & 0 & 0 & 0 
\end{array}
\right)
~,~
Q_{F}^{15b}=
\left(
\begin{array}{cccc|ccccc}
p_1 & p_2 & p_3 & p_4 & s_1 & s_2 & s_3 & s_4 & s_5 \\
\hline
 1 & 1 & 0 & 0 & 0 & 0 & -1 & -1 & 0 \\
 0 & 0 & 1 & 1 & 0 & 0 & -1 & 0 & -1 \\
 0 & 0 & 0 & 0 & 1 & 1 & 0 & -1 & -1
\end{array}
\right)~.
$
}
\\

\noindent
From the quiver incidence matrices, one obtains the following D-term charge matrices

\noindent\makebox[\textwidth]{%
\footnotesize
$
Q_{D}^{13}=
\left(
\begin{array}{ccc|cc|cccc}
 p_{1} & p_{2} & p_{3} & q_{1} & q_{2} & s_{1} & s_{2} & s_{3} & s_{4} \\
   \hline
 0 & 0 & 0 & 1 & -1 & 0 & 0 & 0 & 0 \\
 0 & 0 & 0 & 0 & 0 & 1 & -1 & 0 & 0 \\
 0 & 0 & 0 & 0 & 0 & 0 & 0 & 1 & -1 
\end{array}
\right)
~,~
Q_{D}^{15b}=
\left(
\begin{array}{cccc|ccccc}
p_1 & p_2 & p_3 & p_4 & s_1 & s_2 & s_3 & s_4 & s_5 \\
\hline
 0 & 0 & 0 & 0 & 0 & 1 & -1 & 0 & 0 \\
 0 & 0 & 0 & 0 & 0 & 0 & 1 & -1 & 0 \\
 0 & 0 & 0 & 0 & 0 & 0 & 0 & 1 & -1 
\end{array}
\right)~.
$
}
\\

\noindent
The kernel of the total charge matrix $Q_t$ leads to the coordinates of points in the toric diagram,

\noindent\makebox[\textwidth]{%
\footnotesize
$
G_t^{13}=
\left(
\begin{array}{ccc|cc|cccc}
 p_{1} & p_{2} & p_{3} & q_{1} & q_{2} & s_{1} & s_{2} & s_{3} & s_{4} \\
   \hline
 0 & 0 & 2 & 0 & 0 & 1 & 1 & 1 & 1 \\
 2 & 0 & -1 & 1 & 1 & 0 & 0 & 0 & 0 \\
 1 & 1 & 1 & 1 & 1 & 1 & 1 & 1 & 1
\end{array}
\right)
~,~
G_t^{15b}=
\left(
\begin{array}{cccc|ccccc}
p_1 & p_2 & p_3 & p_4 & s_1 & s_2 & s_3 & s_4 & s_5 \\
\hline
 2 & 0 & 2 & 0 & 1 & 1 & 1 & 1 & 1 \\
 0 & 0 & -1 & 1 & 0 & 0 & 0 & 0 & 0 \\
 1 & 1 & 1 & 1 & 1 & 1 & 1 & 1 & 1
\end{array}
\right)~.
$
}
\\

\noindent
Note that the corresponding toric diagrams in \fref{f13} and \fref{f15b} are $GL(2,\mathbb{Z})$ transformed.

The columns in the $G_t$ matrices indicate the coordinates of points in the toric diagram with the associated perfect matchings. Using this information, one relates columns of the matrices $Q_F$, $Q_D$ and $P$ to either external or internal perfect matchings.

Specular duality swaps external and internal perfect matchings as follows
\beal{essx1}
(p_1,p_2,p_3,q_1,q_2,s_1,s_2,s_3,s_4)_{13} \leftrightarrow
(s_1,s_2,s_3,s_4,s_5,p_1,p_2,p_3,p_4)_{15b}~~.
\eea
Accordingly, the duality maps the perfect matching matrix $P^{13}$ to $P^{15b}$ as well as the F-term charge matrix $Q_{F}^{13}$ to $Q_{F}^{15b}$ by a swap of matrix columns. As a result, the following symplectic quotient descriptions of the master spaces $\master$ are isomorphic
\beal{essx2}
\master_{13}&=&
\mathbb{C}^{9}[p_{1},p_{2},p_{3},q_{1},q_{2},s_{1},s_{2},s_{3},s_{4}]//Q_F^{13}
~,~\nn\\
\master_{15b}&=&
\mathbb{C}^{9}[p_1,p_2,p_3,p_4,s_1,s_2,s_3,s_4,s_5]//Q_F^{15b}~.
\eea

Specular duality can therefore be observed on the level of the Hilbert series of $\master$. Starting with Model 15b, its symplectic quotient leads to the following refined Hilbert series 
\beal{essx3}
g_1(t_i,y_{s_i};\master_{15b})
&=&
\prod_{i=1}^{3}\oint_{|z_i|=1}
\frac{\ud z_i}{2\pi i z_i}~
\frac{1}{
(1-z_1 t_1)
(1-z_1 t_2)
(1-z_2 t_3)
(1-z_2 t_4)
(1-z_3 s_1)
}
\nn\\
&&
\hspace{3cm}
\times
\frac{1}{
(1-z_3 s_2)
(1-\frac{1}{z_1 z_2} s_3)
(1-\frac{1}{z_1 z_3} s_4)
(1-\frac{1}{z_2 z_3} s_5)
}
\nn\\
&=&
\frac{P(t_i,y_{s_i})}{
(1-t_1 t_2 y_{s_3})
(1-t_2 t_3 y_{s_3})
(1-t_1 t_4 y_{s_3})
(1-t_2 t_4 y_{s_3})
}
\nn\\
&&
\times
\frac{1}{
(1-t_1 s_1 y_{s_4})
(1-t_2 s_1 y_{s_4})
(1-t_1 y_{s_2} y_{s_4})
(1-t_2 y_{s_2} y_{s_4})
}
\nn\\
&&
\times
\frac{1}{
(1-t_3 y_{s_1} y_{s_5})
(1-t_4 y_{s_1} y_{s_5})
(1-t_3 y_{s_2} y_{s_5})
(1-t_4 y_{s_2} y_{s_5})
}
~~,
\nn\\
\eea
where the numerator $P(t_i,y_{s_i})$ is presented in appendix \sref{appnum1}. Fugacities $t_i$ and $y_{s_i}$ count external and internal perfect matchings $p_i$ and $s_i$ of Model 15b respectively. The plethystic logarithm of the Hilbert series is
\beal{essx4}
&&
PL[g_1(t_i,y_{s_i};\master_{15b})]=
y_{s_1} y_{s_4} t_1 
+ y_{s_2} y_{s_4} t_1
+ y_{s_1} y_{s_4} t_2 
+ y_{s_2} y_{s_4} t_2 
+ y_{s_1} y_{s_5} t_3 
+ y_{s_2} y_{s_5} t_3 
 \nn\\
 &&
 \hspace{0.5cm}
+ y_{s_1} y_{s_5} t_4 
+ y_{s_2} y_{s_5} t_4
+ y_{s_3} t_1 t_3 
+ y_{s_3} t_2 t_3 
+ y_{s_3} t_1 t_4 
+ y_{s_3} t_2 t_4
- y_{s_1} y_{s_2} y_{s_4} y_{s_5} t_1 t_3 
\nn\\
&&
\hspace{0.5cm}
- y_{s_1} y_{s_2} y_{s_4} y_{s_5} t_2 t_3 
- y_{s_1} y_{s_2} y_{s_4} y_{s_5} t_1 t_4
- y_{s_1} y_{s_2} y_{s_4} y_{s_5} t_2 t_4 
- y_{s_1} y_{s_2} y_{s_4}^2 t_1 t_2 
- y_{s_1} y_{s_2} y_{s_5}^2 t_3 t_4 
 \nn\\
 &&
 \hspace{0.5cm}
- y_{s_1} y_{s_3} y_{s_4} t_1 t_2 t_3 
- y_{s_2} y_{s_3} y_{s_4} t_1 t_2 t_3 
- y_{s_1} y_{s_3} y_{s_4} t_1 t_2 t_4 
- y_{s_2} y_{s_3} y_{s_4} t_1 t_2 t_4 
- y_{s_1} y_{s_3} y_{s_5} t_1 t_3 t_4 
 \nn\\
 &&\hspace{0.5cm}
- y_{s_2} y_{s_3} y_{s_5} t_1 t_3 t_4 
- y_{s_1} y_{s_3} y_{s_5} t_2 t_3 t_4 
- y_{s_2} y_{s_3} y_{s_5} t_2 t_3 t_4 
- y_{s_3}^2 t_1 t_2 t_3 t_4
 +\dots~~.
\eea
It is not finite and therefore indicates that the master space is not a complete intersection.

By specular duality, we obtain the Hilbert series in terms of the perfect matching fugacities of Model 13. The perfect matching map in \eref{essx1} translates to the fugacity map 
\beal{essx5}
(y_{s_i},t_{1,2,3},y_{q_{1,2}})_{13} \leftrightarrow
(t_i,y_{s_{1,2,3}},y_{s_{4,5}})_{15b}~~,
\eea
where $(y_{s_i},t_{1,2,3},y_{q_{1,2}})$ are the fugacities for perfect matchings $(s_i,t_{1,2,3},q_{1,2})$ of Model 13 respectively.
\\

\subsection{Global Symmetries and the Hilbert Series \label{s5_2}}

In order to discuss global symmetries, let us introduce the notation of subscripts and superscripts on groups which refer to fugacities and model numbers respectively.

The F-term charge matrix for Model 13 indicates that the global symmetry is $SU(2)_{x}^{[13]} \times U(1)_f^{[13]} \times SU(2)_{h_1}^{[13]} \times SU(2)_{h_2}^{[13]} \times U(1)_{b}^{[13]} \times U(1)_R^{[13]}$, where $SU(2)_{x}^{[13]}\times U(1)_{f}^{[13]} \times U(1)_{R}^{[13]}$ represents the mesonic symmetry, $SU(2)_{h_1}^{[13]}\times SU(2)_{h_2}^{[13]}$ the hidden baryonic symmetry, and $U(1)_{b}^{[13]}$ the remaining baryonic symmetry. In comparison, for Model 15b, where internal and external perfect matchings are swapped under specular duality, the global symmetry is $SU(2)_{x}^{[15b]}\times SU(2)_{y}^{[15b]} \times SU(2)_{h_1}^{[15b]} \times U(1)_{h_2}^{[15b]} \times U(1)_{b}^{[15b]} \times U(1)_R^{[15b]}$. The mesonic symmetry is $SU(2)_{x}^{[15b]}\times SU(2)_{y}^{[15b]} \times U(1)_R^{[15b]}$, the hidden  baryonic symmetry is $SU(2)_{h_1}^{[15b]}\times U(1)_{h_2}^{[15b]}$, and the remaining baryonic symmetry is $U(1)_{b}^{[15b]}$.

Accordingly, we observe that the swap of external and internal perfect matchings under specular duality leads to the following correspondence between global symmetries
\beal{ess2x1}
SU(2)_{x}^{[13]}\times U(1)_{f}^{[13]}
&\leftrightarrow&
SU(2)_{h_1}^{[15b]}\times U(1)_{h_2}^{[15b]} 
\nn\\
SU(2)_{h_1}^{[13]}\times SU(2)_{h_2}^{[13]}
&\leftrightarrow&
SU(2)_{x}^{[15b]}\times SU(2)_{y}^{[15b]} 
\nn\\
U(1)_{b}^{[13]}
&\leftrightarrow&
U(1)_{b}^{[15b]}
~~.
\eea
It is a swap between mesonic flavour and hidden baryonic symmetries.

Following the discussion in appendix \sref{appch}, one can find global charges on perfect matchings such that the swap of external and internal perfect matchings corresponds to a swap of mesonic flavor and hidden baryonic symmetry charges. A choice of such perfect matching charges for Model 13 and Model 15b is in \tref{t13} and \tref{t15b} respectively.

\begin{table}[H]
\centering
\begin{tabular}{|c|c|c|c|c|c|c|l|} 
\hline
\; 
& $SU(2)_{x}$
& $U(1)_{f}$
& $SU(2)_{h_1}$ 
& $SU(2)_{h_2}$ 
& $U(1)_{b}$
& $U(1)_R$ & fugacity \\
\hline
\hline
$p_1$ & +1 &+1 	& 0 &  0  &  0 & 2/3 & $t_1$\\
$p_2$ & -1 &+1 	& 0 &  0  &  0 & 2/3 & $t_2$\\
$p_3$ &  0 &-2	& 0 &  0  &  0 & 2/3 & $t_3$\\
$q_1$ &  0 & 0	& 0 &  0  & +1 &   0 & $y_{q_1}$\\
$q_2$ &  0 & 0 	& 0 &  0  & -1 &   0 & $y_{q_2}$\\
$s_1$ &  0 & 0	&+1 &  0  &  0 &   0 & $y_{s_1}$\\
$s_2$ &  0 & 0	&-1 &  0  &  0 &   0 & $y_{s_2}$\\
$s_3$ &  0 & 0 	& 0 & +1  &  0 &   0 & $y_{s_3}$\\
$s_4$ &  0 & 0 	& 0 & -1  &  0 &   0 & $y_{s_4}$\\
\hline
\end{tabular}
\caption{Perfect matchings of Model 13 with global charge assignment.\label{t13}}
\end{table}

\begin{table}[H]
\centering
\begin{tabular}{|c|c|c|c|c|c|c|l|} 
\hline
\; 
& $SU(2)_{x}$
& $SU(2)_{y}$ 
& $SU(2)_{h_1}$ 
& $U(1)_{h_2}$
& $U(1)_{b}$
& $U(1)_R$ & fugacity \\
\hline
\hline
$p_1$ & +1 &  0 &  0 &  0 &  0 & 1/2 & $t_1$\\
$p_2$ & -1 &  0 &  0 &  0 &  0 & 1/2 & $t_2$\\
$p_3$ &  0 & +1 &  0 &  0 &  0 & 1/2 & $t_3$\\
$p_4$ &  0 & -1 &  0 &  0 &  0 & 1/2 & $t_4$\\
$s_1$ &  0 &  0 & +1 & +1 &  0 &   0 & $y_{s_1}$\\
$s_2$ &  0 &  0 & -1 & +1 &  0 &   0 & $y_{s_2}$\\
$s_3$ &  0 &  0 &  0 & -2 &  0 &   0 & $y_{s_3}$\\
$s_4$ &  0 &  0 &  0 &  0 & +1 &   0 & $y_{s_4}$\\
$s_5$ &  0 &  0 &  0 &  0 & -1 &   0 & $y_{s_5}$\\
\hline
\end{tabular}
\caption{Perfect matchings of Model 15b with global charge assignment.\label{t15b}}
\end{table}

Starting from Model 15b, the following fugacity map
\beal{ess2x2}
&
t = (y_{s_1} y_{s_2} y_{s_3} y_{s_4} y_{s_5} t_1 t_2 t_3 t_4)^{1/4} ~,~
x = t_1^{1/2} t_2^{-1/2} ~,~
y = t_3^{1/2} t_4^{-1/2} ~,~
&
\nn\\
&
b = (y_{s_4} y_{s_5})^{1/2}~ (t_1 t_2)^{1/4}~ (t_3 t_4)^{-1/4} ~,~
h_1 = y_{s_1}^{1/2} y_{s_2}^{-1/2} ~,~
h_2 = (y_{s_1} y_{s_2} y_{s_4} y_{s_5})^{1/4} ~ y_{s_3}^{-1/4} 
~,
&\nn\\
\eea
leads to the refined Hilbert series in \eref{essx3} and the corresponding plethystic logarithm in \eref{essx4} in terms of characters of irreducible representations of the global symmetry. The expansion of the Hilbert series takes the form
\beal{ess2x3}
&&
g_{1}(t,x,y,h_i,b;\master_{15b})
=
\nn\\
&&
\hspace{1cm}
\sum_{n_1=0}^{\infty} \sum_{n_2=0}^{\infty} \sum_{n_3=0}^{\infty}
~
h_2^{n_1+n_2 -2n_3}
b^{- n_1 + n_2}
~
[n_2+n_3; n_1+n_3; n_1+n_2] t^{n_1+n_2+2n_3}~,
\nn\\
\eea
where $[n_1;n_2;n_3]\equiv [n_1]_x [n_2]_y [n_3]_{h_{1}}$ is the combined character of representations of $SU(2)_x \times SU(2)_y \times SU(2)_{h_1}$.\footnote{cf. \cite{Forcella:2008ng} with a choice of charges on fields which relates to the choice presented here. The identification $F_1=SU(2)_x$, $F_2=SU(2)_y$, $A_2=SU(2)_{h_1}$, $A_1=U(1)_{h_2}$, $B=U(1)_{b}$ and $R=U(1)_R$
is made.} The corresponding plethystic logarithm is
\beal{ess2x3b}
PL[g_{1}(t,x,y,h_i,b;\master_{15b})]&=&
[1;0;1] h_2 b t
+ [0;1;1] h_2 b^{-1} t
+ [1;1;0] h_2^{-2} t^2
\nn\\
&&
- [1;1;0] h_2^2 t^2
- [1;0;1] h_2^{-1} b^{-1} t^3
- [0;1;1] h_2^{-1} b t^3
\nn\\
&&
- h_2^2 b^2 t^2
- h_2^2 b^{-2} t^2
- h_2^{-4} t^4
+\dots ~~.
\eea

In comparison, in terms of global charges on perfect matchings of Model 13, the fugacity map
\beal{ess2x4}
&
t = (y_{s_1} y_{s_2} y_{s_3} y_{s_4} y_{q_1} y_{q_2} t_1 t_2 t_3)^{1/3}
~,~
x = t_1^{1/2} t_2^{-1/2}
~,~
&
\nn\\
&
f = (y_{s_1} y_{s_2} y_{s_3} y_{s_4})^{-1/12} ~(y_{q_1} y_{q_2} t_1 t_2)^{1/6} ~t_3^{-1/3}
~,~
&
\nn\\
&
h_1 = y_{s_1}^{1/2} y_{s_2}^{-1/2}
~,~
h_2 = y_{s_3}^{1/2} y_{s_4}^{-1/2}
~,~
&
\nn\\
&
b = (y_{s_1} y_{s_2})^{1/4}~ (y_{s_3} y_{s_4})^{-1/4} ~y_{q_1}^{1/2} y_{q_2}^{-1/2}
~,~&
\eea 
leads to the following Hilbert series
\beal{ess2x4b}
&&
g_{1}(t,x,f,h_i,b;\master_{13})
=
\nn\\
&& \hspace{1cm}
\sum_{n_1=0}^{\infty}
\sum_{n_2=0}^{\infty}
\sum_{n_3=0}^{\infty}
f^{n_1+n_2-2n_3}
b^{-n_1 + n_2}
~
[n_1 + n_2;n_2+n_3;n_1+n_3] 
~
t^{n_1+n_2+n_3}~,
\nn\\
\eea
where $[n_1;n_2;n_3]\equiv [n_1]_x [n_2]_{h_1} [n_3]_{h_2}$ is the combined character of representations of $SU(2)_{x}\times SU(2)_{h_1} \times SU(2)_{h_2}$. 

The $U(1)_R$ charges on perfect matchings of Model 15b are not mapped by specular duality to $U(1)_R$ charges on perfect matchings of Model 13. This is mainly because only extremal perfect matchings carry non-zero R-charges. In order to illustrate specular duality in terms of the refined Hilbert series, one can without loosing track of the algebraic structure of the moduli space mix the $U(1)_R$ symmetry with the remaining symmetry. This effectively modifies the charge assignment under the global symmetry.\footnote{The algebraic structure of the moduli space is not lost when the orthogonality of global charges on perfect matchings is preserved as discussed in appendix \sref{appch}.} The modification is done via the fugacity map
\beal{ess2x5}
&
\tilde{t} = (y_{s_1} y_{s_2} y_{s_3} y_{s_4} y_{q_1} y_{q_2} t_1 t_2 t_3)^{1/4}
~,~
x = t_1^{1/2} t_2^{-1/2}
~,~
&
\nn\\
&
\tilde{f} = (y_{q_1} y_{q_2} t_1 t_2)^{1/4}~ t_3^{-1/4}
~,~
&
\nn\\
&
h_1 = y_{s_1}^{1/2} y_{s_2}^{-1/2}
~,~
h_2 = y_{s_3}^{1/2} y_{s_4}^{-1/2}
~,~
&
\nn\\
&
b = (y_{s_1} y_{s_2})^{1/4}~ (y_{s_3} y_{s_4})^{-1/4}~ y_{q_1}^{1/2} y_{q_2}^{-1/2}
~,&
\eea
which leads to the Hilbert series
\beal{ess2x6}
&&
g_1(\tilde{t},x,f,h_i,b;\master_{13})
=
\nn\\
&&
\hspace{1cm}
\sum_{n_1=0}^{\infty} 
\sum_{n_2=0}^{\infty} 
\sum_{n_3=0}^{\infty} 
\tilde{f}^{n_1+n_2-2n_3} b^{-n_1 + n_2} [n_1 + n_2; n_2 + n_3; n_1 + n_3] \tilde{t}^{n_1+n_2+2n_3}
~~,\nn\\
\eea
where $[n_1;n_2;n_3]\equiv [n_1]_x [n_2]_{h_1} [n_3]_{h_2}$.
One observes that the fugacity map equivalent to the exchange of mesonic flavour and hidden baryonic symmetries is
\beal{ess2x7}
(x,\tilde{f},\tilde{t},h_1,h_2,b)_{13} 
\leftrightarrow
(h_1,h_2,t,x,y,b)_{15b}~~.
\eea
It relates the Hilbert series in \eref{ess2x3} to the one in \eref{ess2x6}.
\\

\subsection{Generators, the Master Space Cone and the Hilbert Series \label{s5_2}}

The master space is toric Calabi-Yau and has a conical structure. Since the dimension of the master space is $G+2=6$, the corresponding Hilbert series can be rewritten in terms of $6$ fugacities $T_i$ such that the exponents of $T_i$ are positive only. This means that all elements of the ring and the corresponding integral points of the moduli space cone relate to monomials of the form $\prod_i T_i^{m_i}$ with $m_i\geq 0$ in the Hilbert series expansion. The appropriate interpretation for these monomials is that if $b$ $T_i$ vanish in $\prod_i T_i^{m_i}$, the associated integral point is on a codimension $b$ cone. All points associated to monomials $\prod_i T_i^{m_i}$ with $m_i>0$ for all $i$ lie within the codimension $0$ cone. The boundary of the codimension $0$ cone is defined by monomials of the form $T_i^{m_i}$ with $m_i>0$.

\begin{figure}[ht!]
\begin{center}
\includegraphics[width=8 cm]{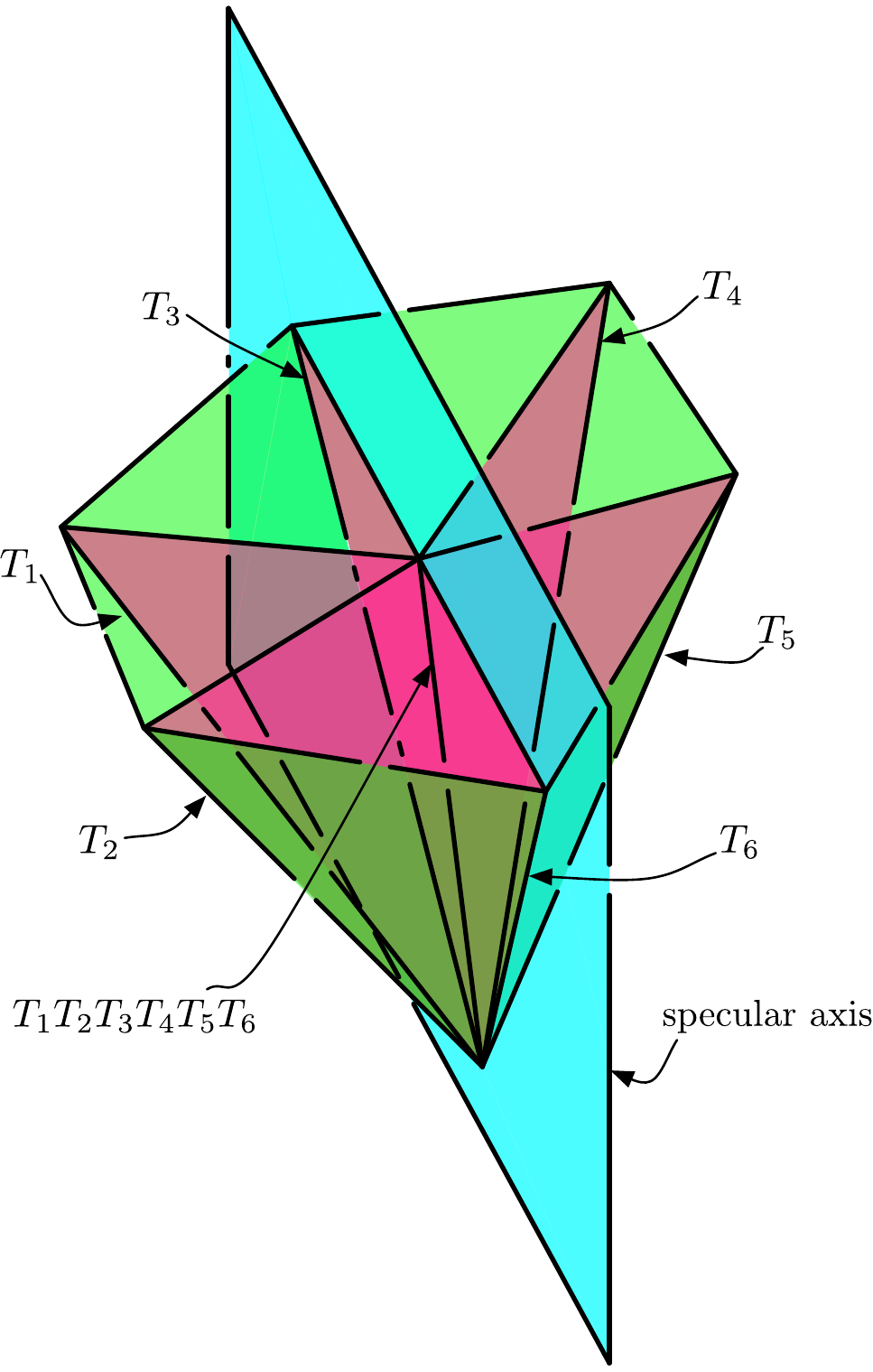}
\caption{\textit{The Specular Axis.} This is a schematic illustration of the master space cone of Models 13 and 15b. The rays corresponding to the basis of the cone are labelled with the associated fugacities $T_i$ of the Hilbert series. The cone is symmetric along a hyperplane which we call the specular axis.}
  \label{fspec1315b}
 \end{center}
 \end{figure}

Starting with the perfect matchings of Model 15b, the fugacity map
\beal{ess3x1}
&
T_1 = x = t_1^{1/2} t_2^{-1/2}~,~
T_2 = y = t_3^{1/2} t_4^{-1/2}~,~
&
\nn\\
&
T_3 = b = (y_{s_4} y_{s_5})^{1/2}~ (t_1 t_2)^{1/4}~ (t_3 t_4)^{-1/4}~,~
&
\nn\\
&
T_4 = h_1 = y_{s_1}^{1/2} y_{s_2}^{-1/2}~,~
T_5 = h_2 = (y_{s_1} y_{s_2} y_{s_5})^{1/4} y_{s_3}^{-1/4} ~,~
&
\nn\\
&
T_6 = \frac{t}{x y b h_1 h_2} =
(y_{s_1} y_{s_2} y_{s_3} y_{s_4} y_{s_5} t_1 t_2 t_3 t_4)^{1/4}
~,~
&
\eea
allows us to re-write the Hilbert series such that the corresponding plethystic logarithm in \eref{essx4} takes the form
\beal{ess3x2}
&&
PL[g(T_i;\master_{15b})] =
T_1^2 T_2 T_3^2 T_4^2 T_5^2 T_6
+T_1^2 T_2 T_3^2 T_5^2 T_6
+T_2 T_3^2 T_4^2 T_5^2 T_6
+T_2 T_3^2 T_5^2 T_6
\nn\\
&& \hspace{0.5cm}
+T_1 T_2^2 T_4^2 T_5^2 T_6
+T_1 T_2^2 T_5^2 T_6
+T_1 T_4^2 T_5^2 T_6
+T_1 T_5^2 T_6
+T_1^3 T_2^3 T_3^2 T_4^2 T_6^2
+T_1 T_2^3 T_3^2 T_4^2 T_6^2
\nn\\
&& \hspace{0.5cm}
+T_1^3 T_2 T_3^2 T_4^2 T_6^2
+T_1 T_2 T_3^2 T_4^2 T_6^2
-T_1^3 T_2^3 T_3^2 T_4^2 T_5^4 T_6^2
-T_1 T_2^3 T_3^2 T_4^2 T_5^4 T_6^2
\nn\\
&& \hspace{0.5cm}
-T_1^3 T_2 T_3^2 T_4^2 T_5^4 T_6^2
-T_1 T_2 T_3^2 T_4^2 T_5^4 T_6^2
-T_1^2 T_2^2 T_3^4 T_4^2 T_5^4 T_6^2
-T_1^2 T_2^2 T_4^2 T_5^4 T_6^2
\nn\\
&& \hspace{0.5cm}
-T_1^3 T_2^4 T_3^4 T_4^4 T_5^2 T_6^3
-T_1^3 T_2^4 T_3^4 T_4^2 T_5^2 T_6^3
-T_1^3 T_2^2 T_3^4 T_4^4 T_5^2 T_6^3
-T_1^3 T_2^2 T_3^4 T_4^2 T_5^2 T_6^3
\nn\\
&& \hspace{0.5cm}
-T_1^4 T_2^3 T_3^2 T_4^4 T_5^2 T_6^3
-T_1^4 T_2^3 T_3^2 T_4^2 T_5^2 T_6^3
-T_1^2 T_2^3 T_3^2 T_4^4 T_5^2 T_6^3
-T_1^2 T_2^3 T_3^2 T_4^2 T_5^2 T_6^3
\nn\\
&& \hspace{0.5cm}
-T_1^4 T_2^4 T_3^4 T_4^4 T_6^4
+\dots ~~.
\eea
As desired, the plethystic logarithm as for the Hilbert series is such that the exponents of the fugacities $T_i$ are positive. In comparison, in relation to perfect matchings of Model 13, the fugacity map 
\beal{ess3x3}
T_1 = x
~,~
T_2 = \tilde{f}
~,~
T_3 = b
~,~
T_4 = h_1
~,~
T_5 = h_2
~,~
T_6 = \frac{\tilde{t}}{x \tilde{f} b h_1 h_2}
~,~
\eea
rewrites the Hilbert series and plethystic logarithm such that they are related to the ones from Model 15b via 
\beal{ess3x4}
(T_1,T_2,T_3,T_4,T_5,T_6) \leftrightarrow (T_4,T_5,T_3,T_1,T_2,T_6)~~.
\eea
Note that the above map for fugacities $T_i$ relates to the one for global symmetry fugacities in \eref{ess2x7}.

Given that the fugacities $T_i$ relate to the boundary of the Calabi-Yau cone, the above fugacity map can be interpreted as a reflection along a hyperplane which is associated to monomials of the form $T_3^{m_3} T_6^{m_6}$. We call the hyperplane the \textbf{specular axis}. It is schematically illustrated in \fref{fspec1315b}.

\begin{figure}[ht!]
\begin{center}
\includegraphics[width=12 cm]{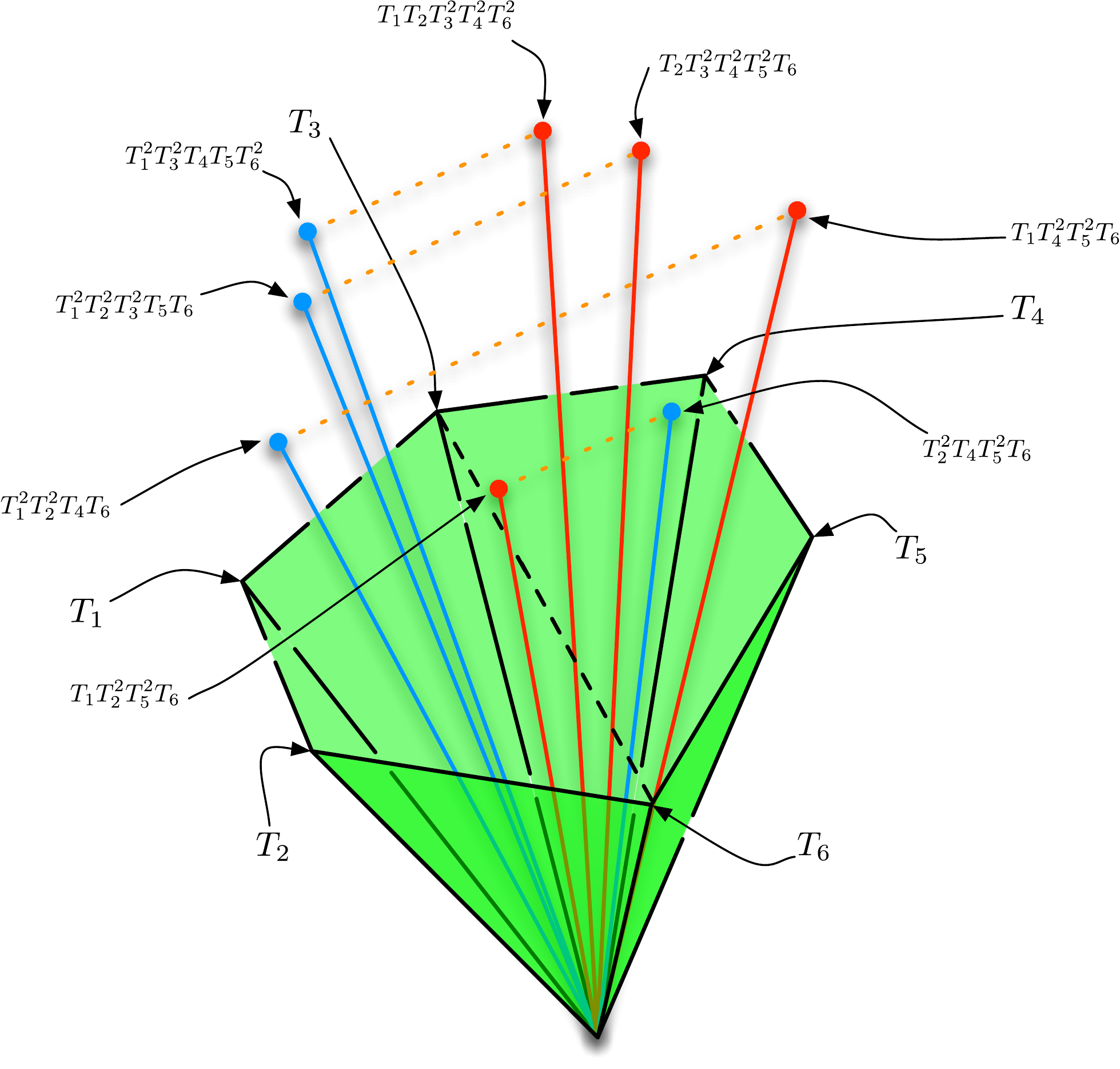}
\caption{\textit{The Specular Axis and Moduli Space Generators.} The schematic illustration shows a selection of master space generators of Model 15b and Model 13 which are highlighted in red and blue respectively. The dotted lines indicate the identifications of generators under specular duality.}
  \label{fspec1315bcone}
 \end{center}
 \end{figure}

The generators of the master space in terms of perfect matchings of Model 13 and Model 15b are shown with the corresponding global symmetry charges in \tref{t13gen} and \tref{t15bgen} respectively. The master space cone with a selection of generators and the specular axis are illustrated schematically in \fref{fspec1315bcone}. Specular duality maps generators into each other along the specular axis.

\begin{table}[H]
\centering
\resizebox{.85\hsize}{!}{
\begin{tabular}{|l|l|c|c|c|c|c|c|l|}
\hline
generator & fields & $SU(2)_x$ & $U(1)_{f}$ &$SU(2)_{h_1}$ & $SU(2)_{h_2}$ &  $U(1)_{b}$ & $U(1)_{R}$ & fugacity
\\
\hline
\hline
$p_3 ~ s_1 s_3$
& $X_{24}$
&  0 & -2
& +1 & +1 &  0 & 1/3
& $T_1^2 T_3^2 T_4^3 T_5^3 T_6^2$
\nn\\
$p_3 ~ s_1 s_4$
& $X_{41}^{1}$
&  0 & -2
& +1 & -1 &  0 & 1/3
& $T_1^2 T_3^2 T_4^3 T_5 T_6^2$
\nn\\
$p_3 ~ s_2 s_3$
& $X_{41}^{1}$
&  0 & -2
& -1 & +1 &  0 & 1/3
& $T_1^2 T_3^2 T_4 T_5^3 T_6^2$
\nn\\
$p_3 ~ s_2 s_4$
& $X_{42}$
&  0 & -2
& -1 & -1 &  0 & 1/3
& $T_1^2 T_3^2 T_4 T_5 T_6^2$
\nn\\
$p_1 ~ q_1 ~ s_1$
& $X_{13}$
& +1 & +1
& +1 &  0 & +1 & 1/3
& $T_1^2 T_2^2 T_3^2 T_4^2 T_5 T_6$
\nn\\
$p_1 ~ q_1 ~ s_2$
& $X_{12}^{2}$
& +1 & +1
& -1 &  0 & +1 & 1/3
& $T_1^2 T_2^2 T_3^2 T_5 T_6$
\nn\\
$p_2 ~ q_1 ~ s_1$
& $X_{34}^{2}$
& -1 & +1
& +1 &  0 & +1 & 1/3
& $T_2^2 T_3^2 T_4^2 T_5 T_6$
\nn\\
$p_2 ~ q_1 ~ s_2$
& $X_{34}^{1}$
& -1 & +1
& -1 &  0 & +1 & 1/3
& $T_2^2 T_3^2 T_5 T_6$
\nn\\
$p_1 ~ q_2 ~ s_3$
& $X_{12}^{1}$
& +1 & +1
&  0 & +1 & -1 & 1/3
& $T_1^2 T_2^2 T_4 T_5^2 T_6$
\nn\\
$p_1 ~ q_2 ~ s_4$
& $X_{31}$
& +1 & +1
&  0 & -1 & -1 & 1/3
& $T_1^2 T_2^2 T_4 T_6$
\nn\\
$p_2 ~ q_2 ~ s_3$
& $X_{23}^{2}$
& -1 & +1
&  0 & +1 & -1 & 1/3
& $T_2^2 T_4 T_5^2 T_6$
\nn\\
$p_2 ~ q_2 ~ s_4$
& $X_{23}^{2}$
& -1 & +1
&  0 & -1 & -1 & 1/3
& $T_2^2 T_4 T_6$
\nn\\
\hline
\end{tabular}
}
\caption{The generators of the master space of Model 13 with the corresponding charges under the global symmetry. \label{t13gen}
}
\end{table}

\begin{table}[H]
\centering
\resizebox{.85\hsize}{!}{
\begin{tabular}{|l|l|c|c|c|c|c|c|l|}
\hline
generator & fields & $SU(2)_x$ & $SU(2)_y$ & $SU(2)_{h_1}$ & $U(1)_{h_2}$ & $U(1)_{b}$ & $U(1)_R$ & fugacity
\\
\hline
\hline
$p_1 p_3 ~ s_3$
& $X_{42}^{2}$
& +1 & +1
&  0 & -2 &  0 & 1
& $T_1^3 T_2^3 T_3^2 T_4^2 T_6^2$
\nn\\
$p_1 p_4 ~ s_3$
& $X_{42}^{4}$
& +1 & -1
&  0 & -2 &  0 & 1
& $T_1^3 T_2 T_3^2 T_4^2 T_6^2$
\nn\\
$p_2 p_3 ~ s_3$
& $X_{42}^{3}$
& -1 & +1
&  0 & -2 &  0 & 1
& $T_1 T_2^3 T_3^2 T_4^2 T_6^2$
\nn\\
$p_2 p_4 ~ s_3$
& $X_{42}^{1}$
& -1 & -1
&  0 & -2 & 0 & 1
& $T_1 T_2 T_3^2 T_4^2 T_6^2$
\nn\\
$p_1 ~ s_1 s_4$
& $X_{21}^{1}$
& +1 & 0
& +1 & +1 & +1 & 1/2
& $T_1^2 T_2 T_3^2 T_4^2 T_5^2 T_6$
\nn\\
$p_2 ~ s_1 s_4$
& $X_{21}^{2}$
& -1 & 0
& +1 & +1 & +1 & 1/2
& $T_2 T_3^2 T_4^2 T_5^2 T_6$
\nn\\
$p_1 ~ s_2 s_4$
& $X_{34}^{2}$
& +1 & 0
& -1 & +1 & +1 & 1/2
& $T_1^2 T_2 T_3^2 T_5^2 T_6$
\nn\\
$p_2 ~ s_2 s_4$
& $X_{34}^{1}$
& -1 & 0
& -1 & +1 & +1 & 1/2
& $T_2 T_3^2 T_5^2 T_6$
\nn\\
$p_3 ~ s_1 s_5$
& $X_{23}^{2}$
& 0 & +1
& +1 &+1 & -1 & 1/2
& $T_1 T_2^2 T_4^2 T_5^2 T_6$
\nn\\
$p_4 ~ s_1 s_5$
& $X_{23}^{1}$
& 0 & -1
& +1 &+1 & -1 & 1/2
& $T_1 T_4^2 T_5^2 T_6$
\nn\\
$p_3 ~ s_2 s_5$
& $X_{14}^{1}$
& 0 & +1
& -1 &+1 & -1 & 1/2
& $T_1 T_2^2 T_5^2 T_6$
\nn\\
$p_4 ~ s_2 s_5$
& $X_{14}^{2}$
& 0 & -1
& -1 &+1 & -1 & 1/2
& $T_1 T_5^2 T_6$
\nn\\
\hline
\end{tabular}
}
\caption{The generators of the master space of Model 15b with the corresponding charges under the global symmetry. \label{t15bgen}
}
\end{table}

\section{Beyond the torus and Conclusions \label{sconc}}

Our work discusses specular duality between brane tilings which represent $4d$ $\mathcal{N}=1$ supersymmetric gauge theories with toric Calabi-Yau moduli spaces. 

Starting from the observations made in \cite{Hanany:2012hi}, this paper identifies the following properties of specular duality for brane tilings on $\mathbb{T}^2$ with reflexive toric diagrams:
\begin{itemize}
\item Dual brane tilings have the same master space $\master$. The corresponding Hilbert series are the same up to a fugacity map.
\item The new correspondence swaps internal and external perfect matchings.
\item Mesonic flavor and anomalous or hidden baryonic symmetries are interchanged.
\item Specular duality represents a hyperplane along which the cone of $\master$ is symmetric.
\end{itemize}
The new duality is an automorphism of the set of $30$ brane tilings with reflexive toric diagrams \cite{Hanany:2012hi}.

When specular duality acts on a brane tiling whose toric diagram is not reflexive, the dual brane tiling is either on a sphere or on a Riemann surface of genus 2 or higher. Such brane tilings have no known AdS duals and their mesonic moduli spaces are not necessarily Calabi-Yau 3-folds \cite{Benvenuti:2004dw,Benvenuti:2005wi,Kennaway:2007tq}. 

In general, the number of faces $G$ of a brane tiling relates to the number of faces $\tilde{G}$ of the dual tiling by
\beal{esco_1}
\tilde{G} = E = G - 2 I+2~.
\eea
$I$ and $E$ are respectively the number of internal and external toric points for the original brane tiling.

\begin{figure}[H]
\begin{center}
\includegraphics[trim=0cm 0cm 0cm 0cm,width=8 cm]{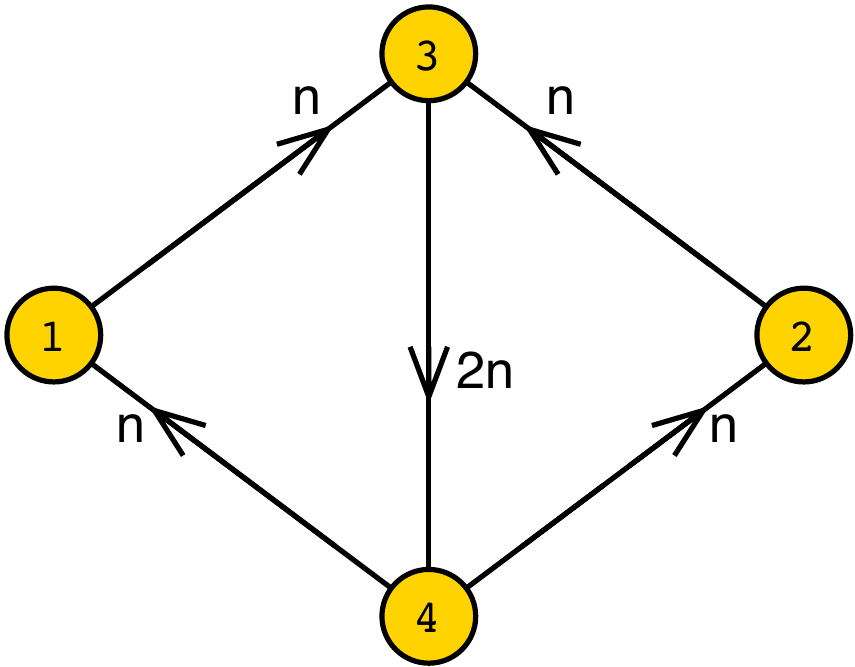}
\caption{The quiver of the specular dual of the brane tiling for the Abelian orbifold of the form $\mathbb{C}^3/\mathbb{Z}_{2n}$ with orbifold action $(1,1,-2)$ \cite{HananySeong11b}.}
  \label{C3Z2quiver}
 \end{center}
 \end{figure}

First examples of brane tilings on Riemann surfaces can be generated from Abelian orbifolds of $\mathbb{C}^3$ \cite{Hanany:2010cx,Davey:2011dd,Hanany:2010ne,Davey:2010px,Hanany:2011iw}. Consider the brane tilings which correspond to the Abelian orbifolds of the form $\mathbb{C}^3/\mathbb{Z}_{2n}$ with orbifold action $(1,1,-2)$ and $n>0$. The dual brane tiling is on a Riemann surface of genus $n-1$. For the first few examples with $n=1,2,3$, the superpotentials are
\beal{esc_1}
W_{\widetilde{\mathbb{C}^3/\mathbb{Z}_{2,(1,1,0)}}}
&=&
X_{34}^{1} X_{41} X_{13}
+ X_{34}^{2} X_{42} X_{23}
- X_{34}^{2} X_{41} X_{13}
- X_{34}^{1} X_{42} X_{23}
~~,
\\
W_{\widetilde{\mathbb{C}^3/\mathbb{Z}_{4,(1,1,2)}}}
&=&
X_{34}^{1} X_{41}^{1} X_{13}^{1}
+ X_{34}^{2} X_{42}^{1} X_{23}^{1}
+ X_{34}^{3} X_{41}^{2} X_{13}^{2}
+ X_{34}^{4} X_{42}^{2} X_{23}^{2}
\nn\\
&&
- X_{34}^{4} X_{41}^{2} X_{13}^{1}
- X_{34}^{1} X_{42}^{2} X_{23}^{1}
- X_{34}^{2} X_{41}^{1} X_{13}^{2}
- X_{34}^{3} X_{42}^{1} X_{23}^{2}
~~,
\\
W_{\widetilde{\mathbb{C}^3/\mathbb{Z}_{6,(1,1,4)}}}
&=&
X_{34}^{1} X_{41}^{1} X_{13}^{1}
+ X_{34}^{2} X_{42}^{1} X_{23}^{1}
+ X_{34}^{3} X_{41}^{2} X_{13}^{2}
+ X_{34}^{4} X_{42}^{2} X_{23}^{2}
\nn\\
&&
+ X_{34}^{5} X_{41}^{3} X_{13}^{3}
+ X_{34}^{6} X_{42}^{3} X_{23}^{3}
- X_{34}^{6} X_{41}^{3} X_{13}^{1}
- X_{34}^{1} X_{42}^{3} X_{23}^{1}
\nn\\
&&
- X_{34}^{2} X_{41}^{1} X_{13}^{2}
- X_{34}^{3} X_{42}^{1} X_{23}^{2}
- X_{34}^{4} X_{41}^{2} X_{13}^{3}
- X_{34}^{5} X_{42}^{2} X_{23}^{3}
\comment{
~~,
\\
W_{\widetilde{\mathbb{C}^3/\mathbb{Z}_{8,(1,1,6)}}}
&=&
X_{34}^{1} X_{41}^{1} X_{13}^{1}
+ X_{34}^{2} X_{42}^{1} X_{23}^{1}
+ X_{34}^{3} X_{41}^{2} X_{13}^{2} 
+ X_{34}^{4} X_{42}^{2} X_{23}^{2}
\nn\\
&&
+ X_{34}^{5} X_{41}^{3} X_{13}^{3}
+ X_{34}^{6} X_{42}^{3} X_{23}^{3}
+ X_{34}^{7} X_{41}^{4} X_{13}^{4}
+ X_{34}^{8} X_{42}^{4} X_{23}^{4}
\nn\\
&&
- X_{34}^{8} X_{41}^{4} X_{13}^{1}
- X_{34}^{1} X_{42}^{4} X_{23}^{1}
- X_{34}^{2} X_{41}^{1} X_{13}^{2}
- X_{34}^{3} X_{42}^{1} X_{23}^{2}
\nn\\
&&
- X_{34}^{4} X_{41}^{2} X_{13}^{3}
- X_{34}^{5} X_{42}^{2} X_{23}^{3}
- X_{34}^{6} X_{41}^{3} X_{13}^{4}
- X_{34}^{7} X_{42}^{3} X_{23}^{4}
}
~~.
\eea

\begin{figure}[H]
\begin{center}
\includegraphics[trim=0cm 0cm 0cm 0cm,width=12 cm]{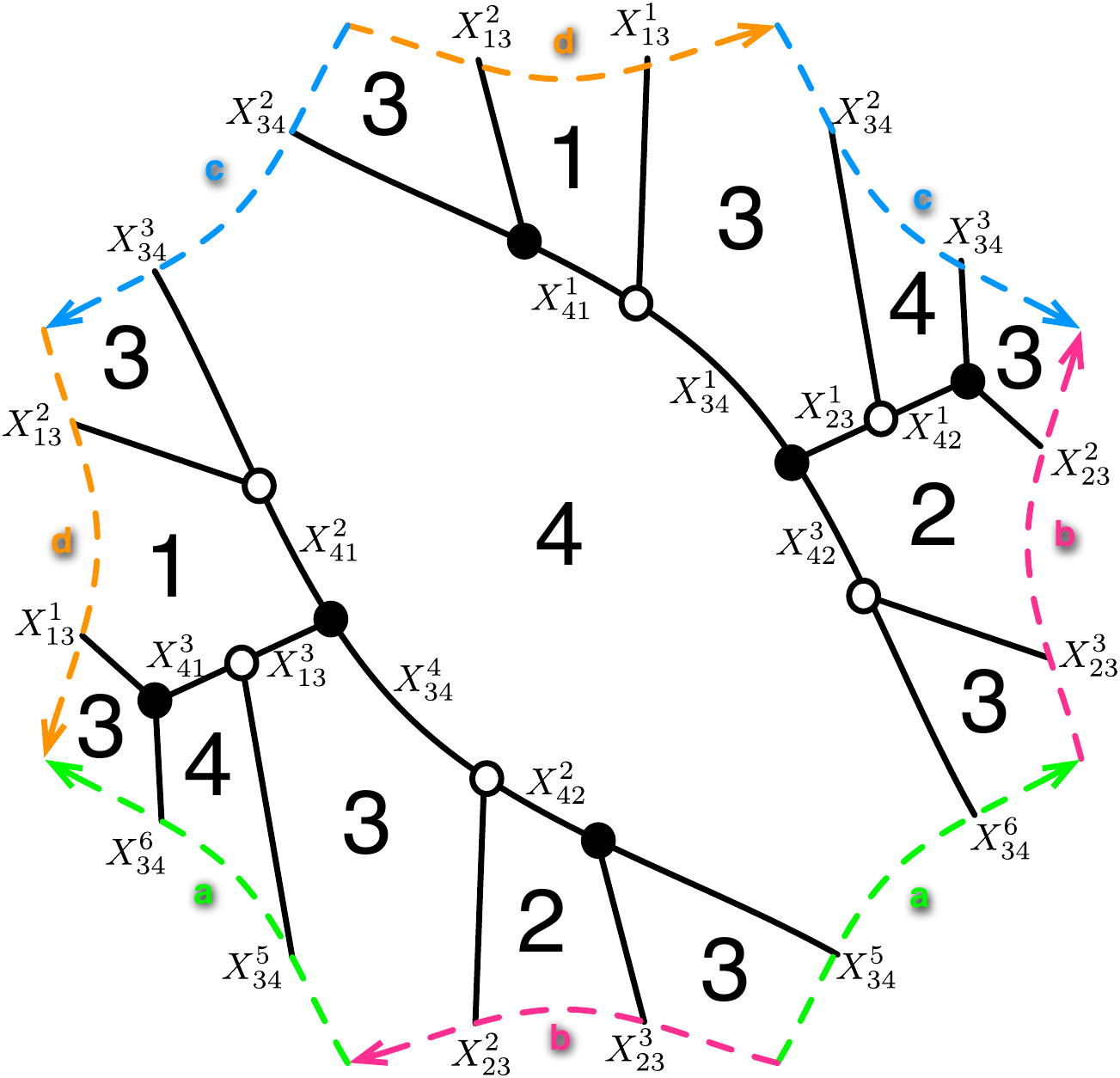}
\caption{\textit{Brane Tiling on a $g=2$ Riemann Surface.} The figure shows the octagonal fundamental domain of the brane tiling which is the specular dual of $\mathbb{C}^3/\mathbb{Z}_6$ with action $(1,1,4)$.}
  \label{fspecc3z6}
 \end{center}
 \end{figure}

The corresponding quivers are shown in \fref{C3Z2quiver}. The Hilbert series of the master spaces are,
\beal{esc_2}
&&
g_{1}(t;\widetilde{\mathbb{C}^3/\mathbb{Z}_{2,(1,1,0)}})
=
\frac{1 - t^4}{(1 - t) (1 - t^2)^4}
~~,
\nn\\
&&
g_{1}(t;\widetilde{\mathbb{C}^3/\mathbb{Z}_{4,(1,1,2)}})
=
\frac{1 + 6 t^3 + 6 t^6 + t^9}{(1 - t^3)^6}
~~,
\nn\\
&&
g_{1}(t;\widetilde{\mathbb{C}^3/\mathbb{Z}_{6,(1,1,4)}})
=
(1 + 3 t^2 + 7 t^4 + 18 t^6 + 38 t^8 + 72 t^{10} + 122 t^{12} + 186 t^{14} + 267 t^{16}
\nn\\
&&
\hspace{2cm}
 + 363 t^{18} + 456 t^{20} + 537 t^{22} + 588 t^{24} + 603 t^{26} + 588 t^{28} + 537 t^{30} + 456 t^{32}
\nn\\
&&
\hspace{2cm}
  + 363 t^{34} + 267 t^{36} + 186 t^{38} + 122 t^{40} + 72 t^{42} + 38 t^{44} + 18 t^{46} + 7 t^{48}
\nn\\
&&
\hspace{2cm}
+ 3 t^{50} + t^{52})
\times
\frac{
(1 - t^2)^3 (1 - t^4) 
}{
(1 - t^6)^7 (1 - t^8)^5  
}~~.
\eea

The fundamental domain of the brane tiling for the specular dual of $\mathbb{C}^3/\mathbb{Z}_{6,(1,1,4)}$ is in \fref{fspecc3z6}. It is of great interest to study such brane tilings on higher genus Riemann surfaces. One obtains a new class of quivers and field theories via specular duality which is the subject of a future investigation \cite{HananySeong11b}.
\\
 
\section*{Acknowledgements}

We like to thank S. Cremonesi, S. Franco and G. Torri for fruitful discussions. We also thank J. Stienstra for interesting correspondence. A. H. thanks Stanford University and SLAC for the kind hospitality during various stages of this project. R.-K. S. is grateful to the Simons Center for Geometry and Physics at Stony Brook University and the Hebrew University of Jerusalem for kind hospitality. 

\appendix

\section{Comments on mesonic and baryonic symmetry charges \label{appch}}

\noindent\textbf{Mesonic Symmetry.} The mesonic moduli space of a given brane tiling on $\mathbb{T}^2$ is a non-compact toric Calabi-Yau $3$-fold. The mesonic symmetry of the quiver gauge theory has rank $3$ and hence takes one of the following forms,
\begin{itemize}
\item $U(1) \times U(1) \times U(1)$
\item $SU(2) \times U(1) \times U(1)$
\item $SU(2) \times SU(2) \times U(1)$
\item $SU(3) \times U(1)$~~~,
\end{itemize}
where the R-symmetry is a subgroup. For $\mathcal{N}=2$ and $\mathcal{N}=1$, the R-symmetry is respectively $SU(2)\times U(1)$ and $U(1)$.

The above global symmetries derive from the isometry group of the Calabi-Yau 3-fold. The enhancement of a $U(1)$ flavor to $SU(2)$ or $SU(3)$ is indicated by columns in the total charge matrix $Q_t$ which carry the same charge and correspond to external perfect matchings.
\\

\noindent\textbf{Baryonic Symmetry.}  The baryonic symmetry is $U(1)^{G-1}$ or an enhancement with rank $G-1$. It is divided into an anomalous and non-anomalous part. The anomalous $U(1)$ baryonic symmetries can appear as enhanced non-Abelian symmetries which are known as hidden symmetries. These are isometries of the master space, but are not a symmetry of the Lagrangian. They are indicated by non-extremal perfect matchings which carry the same $Q_F$ charge. The number of anomalous $U(1)$ baryonic symmetries or the rank of the hidden symmetry is given by twice the number of internal points in the associated toric diagram, $2 I$. The non-anomalous baryonic $U(1)$ symmetries are given by $E-3$ where $E$ is the number of external toric points.
\\

\noindent\textbf{Mesonic and Baryonic Charges on perfect matchings.} The perfect matchings carry $G+2$ charges which relate to the $3$ mesonic and $G-1$ baryonic symmetries. Each perfect matching is assigned a $G+2$ dimensional charge vector, and the choice of its components is arbitrary up to the following constraints:
\begin{itemize}
\item All $G+2$ dimensional charge vectors are linearly independent to each other.
\item The sum of all charge vectors is $(0,\dots,0,2)$ where the non-zero component $2$ is the total $U(1)_R$-charge.
\end{itemize}
Note that if two charge vectors are linearly dependent, information about the algebraic structure of the moduli space is lost. For the purpose of studying specular duality, the following additional constraints are introduced without loosing track of the algebraic structure of the master space:
\begin{itemize}
\item For a pair of dual brane tilings, the charge vectors can be chosen such that a swap between internal and external perfect matchings equates to a swap of mesonic flavour and anomalous or hidden baryonic symmetry charges.
\item If the $U(1)_R$-charges are irrational or otherwise incompatible between dual brane tilings, one can find a set of orthogonal replacement charges without loosing information on the algebraic structure of the master space. This modification corresponds to a mix of the R-symmetry with the remaining global symmetry.
\end{itemize}

\section{Hilbert series of $\master$ for Models 13 and 15b \label{appnum}\label{appnum1}}

The refined Hilbert series of the master space of Model 15b, and by specular duality of the master space of Model 13, is of the form
\beal{essx3app}
g_1(t_i,y_{s_i};\master_{15b})
&=&
\frac{P(t_i,y_{s_i})}{
(1-t_1 t_2 y_{s_3})
(1-t_2 t_3 y_{s_3})
(1-t_1 t_4 y_{s_3})
(1-t_2 t_4 y_{s_3})
}
\nn\\
&&
\times
\frac{1}{
(1-t_1 s_1 y_{s_4})
(1-t_2 s_1 y_{s_4})
(1-t_1 y_{s_2} y_{s_4})
(1-t_2 y_{s_2} y_{s_4})
}
\nn\\
&&
\times
\frac{1}{
(1-t_3 y_{s_1} y_{s_5})
(1-t_4 y_{s_1} y_{s_5})
(1-t_3 y_{s_2} y_{s_5})
(1-t_4 y_{s_2} y_{s_5})
}
~~,
\nn\\
\eea
where the numerator is
\beal{esx3app2}
&&
P(t_i,s_i)
=
\nn\\
&&
1 - t_1 t_2 t_3 t_4 y_{s_3}^2 - t_1 t_2 t_3 y_{s_1} y_{s_3} y_{s_4} - t_1 t_2 t_4 y_{s_1} y_{s_3} y_{s_4} - 
 t_1 t_2 t_3 y_{s_2} y_{s_3} y_{s_4} - t_1 t_2 t_4 y_{s_2} y_{s_3} y_{s_4}
 \nn\\
 && 
 + t_1^2 t_2 t_3 t_4 y_{s_1} y_{s_3}^2 y_{s_4} + 
 t_1 t_2^2 t_3 t_4 y_{s_1} y_{s_3}^2 y_{s_4} + t_1^2 t_2 t_3 t_4 y_{s_2} y_{s_3}^2 y_{s_4} + 
 t_1 t_2^2 t_3 t_4 y_{s_2} y_{s_3}^2 y_{s_4} - t_1 t_2 y_{s_1} y_{s_2} y_{s_4}^2 
 \nn\\
 &&
 + 
 t_1^2 t_2 t_3 y_{s_1} y_{s_2} y_{s_3} y_{s_4}^2 + t_1 t_2^2 t_3 y_{s_1} y_{s_2} y_{s_3} y_{s_4}^2 + 
 t_1^2 t_2 t_4 y_{s_1} y_{s_2} y_{s_3} y_{s_4}^2 + t_1 t_2^2 t_4 y_{s_1} y_{s_2} y_{s_3} y_{s_4}^2 - 
 t_1^3 t_2 t_3 t_4 y_{s_1} y_{s_2} y_{s_3}^2 y_{s_4}^2 
 \nn\\
 &&
 - t_1^2 t_2^2 t_3 t_4 y_{s_1} y_{s_2} y_{s_3}^2 y_{s_4}^2 - 
 t_1 t_2^3 t_3 t_4 y_{s_1} y_{s_2} y_{s_3}^2 y_{s_4}^2 - t_1 t_3 t_4 y_{s_1} y_{s_3} y_{s_5} - 
 t_2 t_3 t_4 y_{s_1} y_{s_3} y_{s_5} - t_1 t_3 t_4 y_{s_2} y_{s_3} y_{s_5} 
 \nn\\
 &&
 - t_2 t_3 t_4 y_{s_2} y_{s_3} y_{s_5} + 
 t_1 t_2 t_3^2 t_4 y_{s_1} y_{s_3}^2 y_{s_5} + t_1 t_2 t_3 t_4^2 y_{s_1} y_{s_3}^2 y_{s_5} + 
 t_1 t_2 t_3^2 t_4 y_{s_2} y_{s_3}^2 y_{s_5} + t_1 t_2 t_3 t_4^2 y_{s_2} y_{s_3}^2 y_{s_5} 
 \nn\\
 &&
 - 
 t_1 t_3 y_{s_1} y_{s_2} y_{s_4} y_{s_5} - t_2 t_3 y_{s_1} y_{s_2} y_{s_4} y_{s_5} - t_1 t_4 y_{s_1} y_{s_2} y_{s_4} y_{s_5} - 
 t_2 t_4 y_{s_1} y_{s_2} y_{s_4} y_{s_5} + t_1 t_2 t_3 t_4 y_{s_1}^2 y_{s_3} y_{s_4} y_{s_5} 
 \nn\\
 &&
 + 
 t_1 t_2 t_3^2 y_{s_1} y_{s_2} y_{s_3} y_{s_4} y_{s_5} + t_1^2 t_3 t_4 y_{s_1} y_{s_2} y_{s_3} y_{s_4} y_{s_5} + 
 5 t_1 t_2 t_3 t_4 y_{s_1} y_{s_2} y_{s_3} y_{s_4} y_{s_5} + t_2^2 t_3 t_4 y_{s_1} y_{s_2} y_{s_3} y_{s_4} y_{s_5} 
 \nn\\
 &&
 + 
 t_1 t_2 t_4^2 y_{s_1} y_{s_2} y_{s_3} y_{s_4} y_{s_5} 
 + t_1 t_2 t_3 t_4 y_{s_2}^2 y_{s_3} y_{s_4} y_{s_5} - 
 t_1^2 t_2 t_3^2 t_4 y_{s_1} y_{s_2} y_{s_3}^2 y_{s_4} y_{s_5} - t_1 t_2^2 t_3^2 t_4 y_{s_1} y_{s_2} y_{s_3}^2 y_{s_4} y_{s_5} 
 \nn\\
 &&
 -
  t_1^2 t_2 t_3 t_4^2 y_{s_1} y_{s_2} y_{s_3}^2 y_{s_4} y_{s_5} 
 - 
 t_1 t_2^2 t_3 t_4^2 y_{s_1} y_{s_2} y_{s_3}^2 y_{s_4} y_{s_5} 
 - 
 t_1^2 t_2^2 t_3^2 t_4^2 y_{s_1}^2 y_{s_3}^3 y_{s_4} y_{s_5} - 
 t_1^2 t_2^2 t_3^2 t_4^2 y_{s_1} y_{s_2} y_{s_3}^3 y_{s_4} y_{s_5} 
 \nn\\
 &&
 - 
 t_1^2 t_2^2 t_3^2 t_4^2 y_{s_2}^2 y_{s_3}^3 y_{s_4} y_{s_5} + t_1 t_2 t_3 y_{s_1}^2 y_{s_2} y_{s_4}^2 y_{s_5} + 
 t_1 t_2 t_4 y_{s_1}^2 y_{s_2} y_{s_4}^2 y_{s_5} 
 + t_1 t_2 t_3 y_{s_1} y_{s_2}^2 y_{s_4}^2 y_{s_5} 
 \nn\\
 &&
 + 
 t_1 t_2 t_4 y_{s_1} y_{s_2}^2 y_{s_4}^2 y_{s_5} - t_1^2 t_2 t_3 t_4 y_{s_1}^2 y_{s_2} y_{s_3} y_{s_4}^2 y_{s_5} - 
 t_1 t_2^2 t_3 t_4 y_{s_1}^2 y_{s_2} y_{s_3} y_{s_4}^2 y_{s_5} - t_1^2 t_2 t_3 t_4 y_{s_1} y_{s_2}^2 y_{s_3} y_{s_4}^2 y_{s_5} 
 \nn\\
 &&
 -
  t_1 t_2^2 t_3 t_4 y_{s_1} y_{s_2}^2 y_{s_3} y_{s_4}^2 y_{s_5} - 
 t_1^2 t_2^2 t_3^2 t_4 y_{s_1}^2 y_{s_2} y_{s_3}^2 y_{s_4}^2 y_{s_5} - 
 t_1^2 t_2^2 t_3 t_4^2 y_{s_1}^2 y_{s_2} y_{s_3}^2 y_{s_4}^2 y_{s_5} - 
 t_1^2 t_2^2 t_3^2 t_4 y_{s_1} y_{s_2}^2 y_{s_3}^2 y_{s_4}^2 y_{s_5} 
 \nn\\
 &&
 - 
 t_1^2 t_2^2 t_3 t_4^2 y_{s_1} y_{s_2}^2 y_{s_3}^2 y_{s_4}^2 y_{s_5} + 
 t_1^3 t_2^2 t_3^2 t_4^2 y_{s_1}^2 y_{s_2} y_{s_3}^3 y_{s_4}^2 y_{s_5} + 
 t_1^2 t_2^3 t_3^2 t_4^2 y_{s_1}^2 y_{s_2} y_{s_3}^3 y_{s_4}^2 y_{s_5} + 
 t_1^3 t_2^2 t_3^2 t_4^2 y_{s_1} y_{s_2}^2 y_{s_3}^3 y_{s_4}^2 y_{s_5} 
 \nn\\
  &&
 + 
 t_1^2 t_2^3 t_3^2 t_4^2 y_{s_1} y_{s_2}^2 y_{s_3}^3 y_{s_4}^2 y_{s_5} - 
 t_1^2 t_2^2 t_3^2 y_{s_1}^2 y_{s_2}^2 y_{s_3} y_{s_4}^3 y_{s_5} - 
 t_1^2 t_2^2 t_3 t_4 y_{s_1}^2 y_{s_2}^2 y_{s_3} y_{s_4}^3 y_{s_5} - 
 t_1^2 t_2^2 t_4^2 y_{s_1}^2 y_{s_2}^2 y_{s_3} y_{s_4}^3 y_{s_5} 
 \nn\\
 &&
 + 
 t_1^3 t_2^2 t_3^2 t_4 y_{s_1}^2 y_{s_2}^2 y_{s_3}^2 y_{s_4}^3 y_{s_5} + 
 t_1^2 t_2^3 t_3^2 t_4 y_{s_1}^2 y_{s_2}^2 y_{s_3}^2 y_{s_4}^3 y_{s_5} + 
 t_1^3 t_2^2 t_3 t_4^2 y_{s_1}^2 y_{s_2}^2 y_{s_3}^2 y_{s_4}^3 y_{s_5} + 
 t_1^2 t_2^3 t_3 t_4^2 y_{s_1}^2 y_{s_2}^2 y_{s_3}^2 y_{s_4}^3 y_{s_5} 
 \nn\\
 &&
 - 
 t_1^3 t_2^3 t_3^2 t_4^2 y_{s_1}^2 y_{s_2}^2 y_{s_3}^3 y_{s_4}^3 y_{s_5} - t_3 t_4 y_{s_1} y_{s_2} y_{s_5}^2 + 
 t_1 t_3^2 t_4 y_{s_1} y_{s_2} y_{s_3} y_{s_5}^2 + t_2 t_3^2 t_4 y_{s_1} y_{s_2} y_{s_3} y_{s_5}^2 
 \nn\\
 &&
 + 
 t_1 t_3 t_4^2 y_{s_1} y_{s_2} y_{s_3} y_{s_5}^2 + t_2 t_3 t_4^2 y_{s_1} y_{s_2} y_{s_3} y_{s_5}^2 - 
 t_1 t_2 t_3^3 t_4 y_{s_1} y_{s_2} y_{s_3}^2 y_{s_5}^2 - t_1 t_2 t_3^2 t_4^2 y_{s_1} y_{s_2} y_{s_3}^2 y_{s_5}^2 
 \nn\\
 &&
 - 
 t_1 t_2 t_3 t_4^3 y_{s_1} y_{s_2} y_{s_3}^2 y_{s_5}^2 + t_1 t_3 t_4 y_{s_1}^2 y_{s_2} y_{s_4} y_{s_5}^2 + 
 t_2 t_3 t_4 y_{s_1}^2 y_{s_2} y_{s_4} y_{s_5}^2 + t_1 t_3 t_4 y_{s_1} y_{s_2}^2 y_{s_4} y_{s_5}^2 
 \nn\\
 &&
 + 
 t_2 t_3 t_4 y_{s_1} y_{s_2}^2 y_{s_4} y_{s_5}^2 - t_1 t_2 t_3^2 t_4 y_{s_1}^2 y_{s_2} y_{s_3} y_{s_4} y_{s_5}^2 - 
 t_1 t_2 t_3 t_4^2 y_{s_1}^2 y_{s_2} y_{s_3} y_{s_4} y_{s_5}^2 - t_1 t_2 t_3^2 t_4 y_{s_1} y_{s_2}^2 y_{s_3} y_{s_4} y_{s_5}^2 
 \nn\\
 &&
 -
  t_1 t_2 t_3 t_4^2 y_{s_1} y_{s_2}^2 y_{s_3} y_{s_4} y_{s_5}^2 - 
 t_1^2 t_2 t_3^2 t_4^2 y_{s_1}^2 y_{s_2} y_{s_3}^2 y_{s_4} y_{s_5}^2 - 
 t_1 t_2^2 t_3^2 t_4^2 y_{s_1}^2 y_{s_2} y_{s_3}^2 y_{s_4} y_{s_5}^2 - 
 t_1^2 t_2 t_3^2 t_4^2 y_{s_1} y_{s_2}^2 y_{s_3}^2 y_{s_4} y_{s_5}^2 
 \nn
 \eea
 \beal{esx3app2b}
 &&
 - 
 t_1 t_2^2 t_3^2 t_4^2 y_{s_1} y_{s_2}^2 y_{s_3}^2 y_{s_4} y_{s_5}^2 + 
 t_1^2 t_2^2 t_3^3 t_4^2 y_{s_1}^2 y_{s_2} y_{s_3}^3 y_{s_4} y_{s_5}^2 + 
 t_1^2 t_2^2 t_3^2 t_4^3 y_{s_1}^2 y_{s_2} y_{s_3}^3 y_{s_4} y_{s_5}^2 + 
 t_1^2 t_2^2 t_3^3 t_4^2 y_{s_1} y_{s_2}^2 y_{s_3}^3 y_{s_4} y_{s_5}^2 
 \nn\\
 &&
 + 
 t_1^2 t_2^2 t_3^2 t_4^3 y_{s_1} y_{s_2}^2 y_{s_3}^3 y_{s_4} y_{s_5}^2 - 
 t_1 t_2 t_3 t_4 y_{s_1}^3 y_{s_2} y_{s_4}^2 y_{s_5}^2 - t_1 t_2 t_3 t_4 y_{s_1}^2 y_{s_2}^2 y_{s_4}^2 y_{s_5}^2 - 
 t_1 t_2 t_3 t_4 y_{s_1} y_{s_2}^3 y_{s_4}^2 y_{s_5}^2 
 \nn\\
 &&
 - 
 t_1^2 t_2 t_3^2 t_4 y_{s_1}^2 y_{s_2}^2 y_{s_3} y_{s_4}^2 y_{s_5}^2 - 
 t_1 t_2^2 t_3^2 t_4 y_{s_1}^2 y_{s_2}^2 y_{s_3} y_{s_4}^2 y_{s_5}^2 - 
 t_1^2 t_2 t_3 t_4^2 y_{s_1}^2 y_{s_2}^2 y_{s_3} y_{s_4}^2 y_{s_5}^2 - 
 t_1 t_2^2 t_3 t_4^2 y_{s_1}^2 y_{s_2}^2 y_{s_3} y_{s_4}^2 y_{s_5}^2 
 \nn\\
 &&
 + 
 t_1^2 t_2^2 t_3^2 t_4^2 y_{s_1}^3 y_{s_2} y_{s_3}^2 y_{s_4}^2 y_{s_5}^2 + 
 t_1^2 t_2^2 t_3^3 t_4 y_{s_1}^2 y_{s_2}^2 y_{s_3}^2 y_{s_4}^2 y_{s_5}^2 + 
 t_1^3 t_2 t_3^2 t_4^2 y_{s_1}^2 y_{s_2}^2 y_{s_3}^2 y_{s_4}^2 y_{s_5}^2 + 
 5 t_1^2 t_2^2 t_3^2 t_4^2 y_{s_1}^2 y_{s_2}^2 y_{s_3}^2 y_{s_4}^2 y_{s_5}^2 
 \nn\\
 &&
 + 
 t_1 t_2^3 t_3^2 t_4^2 y_{s_1}^2 y_{s_2}^2 y_{s_3}^2 y_{s_4}^2 y_{s_5}^2 + 
 t_1^2 t_2^2 t_3 t_4^3 y_{s_1}^2 y_{s_2}^2 y_{s_3}^2 y_{s_4}^2 y_{s_5}^2 + 
 t_1^2 t_2^2 t_3^2 t_4^2 y_{s_1} y_{s_2}^3 y_{s_3}^2 y_{s_4}^2 y_{s_5}^2 - 
 t_1^3 t_2^2 t_3^3 t_4^2 y_{s_1}^2 y_{s_2}^2 y_{s_3}^3 y_{s_4}^2 y_{s_5}^2 
 \nn\\
 &&
 - 
 t_1^2 t_2^3 t_3^3 t_4^2 y_{s_1}^2 y_{s_2}^2 y_{s_3}^3 y_{s_4}^2 y_{s_5}^2 - 
 t_1^3 t_2^2 t_3^2 t_4^3 y_{s_1}^2 y_{s_2}^2 y_{s_3}^3 y_{s_4}^2 y_{s_5}^2 - 
 t_1^2 t_2^3 t_3^2 t_4^3 y_{s_1}^2 y_{s_2}^2 y_{s_3}^3 y_{s_4}^2 y_{s_5}^2 + 
 t_1^2 t_2^2 t_3^2 t_4 y_{s_1}^3 y_{s_2}^2 y_{s_3} y_{s_4}^3 y_{s_5}^2 
 \nn\\
 &&
 + 
 t_1^2 t_2^2 t_3 t_4^2 y_{s_1}^3 y_{s_2}^2 y_{s_3} y_{s_4}^3 y_{s_5}^2 + 
 t_1^2 t_2^2 t_3^2 t_4 y_{s_1}^2 y_{s_2}^3 y_{s_3} y_{s_4}^3 y_{s_5}^2 + 
 t_1^2 t_2^2 t_3 t_4^2 y_{s_1}^2 y_{s_2}^3 y_{s_3} y_{s_4}^3 y_{s_5}^2 - 
 t_1^3 t_2^2 t_3^2 t_4^2 y_{s_1}^3 y_{s_2}^2 y_{s_3}^2 y_{s_4}^3 y_{s_5}^2 
 \nn\\
 &&
 - 
 t_1^2 t_2^3 t_3^2 t_4^2 y_{s_1}^3 y_{s_2}^2 y_{s_3}^2 y_{s_4}^3 y_{s_5}^2 - 
 t_1^3 t_2^2 t_3^2 t_4^2 y_{s_1}^2 y_{s_2}^3 y_{s_3}^2 y_{s_4}^3 y_{s_5}^2 - 
 t_1^2 t_2^3 t_3^2 t_4^2 y_{s_1}^2 y_{s_2}^3 y_{s_3}^2 y_{s_4}^3 y_{s_5}^2 - 
 t_1^2 t_3^2 t_4^2 y_{s_1}^2 y_{s_2}^2 y_{s_3} y_{s_4} y_{s_5}^3 
 \nn\\
 &&
 - 
 t_1 t_2 t_3^2 t_4^2 y_{s_1}^2 y_{s_2}^2 y_{s_3} y_{s_4} y_{s_5}^3 - 
 t_2^2 t_3^2 t_4^2 y_{s_1}^2 y_{s_2}^2 y_{s_3} y_{s_4} y_{s_5}^3 + 
 t_1^2 t_2 t_3^3 t_4^2 y_{s_1}^2 y_{s_2}^2 y_{s_3}^2 y_{s_4} y_{s_5}^3 + 
 t_1 t_2^2 t_3^3 t_4^2 y_{s_1}^2 y_{s_2}^2 y_{s_3}^2 y_{s_4} y_{s_5}^3 
 \nn\\
 &&
 + 
 t_1^2 t_2 t_3^2 t_4^3 y_{s_1}^2 y_{s_2}^2 y_{s_3}^2 y_{s_4} y_{s_5}^3 + 
 t_1 t_2^2 t_3^2 t_4^3 y_{s_1}^2 y_{s_2}^2 y_{s_3}^2 y_{s_4} y_{s_5}^3 - 
 t_1^2 t_2^2 t_3^3 t_4^3 y_{s_1}^2 y_{s_2}^2 y_{s_3}^3 y_{s_4} y_{s_5}^3 + 
 t_1^2 t_2 t_3^2 t_4^2 y_{s_1}^3 y_{s_2}^2 y_{s_3} y_{s_4}^2 y_{s_5}^3 
 \nn\\
 &&
 + 
 t_1 t_2^2 t_3^2 t_4^2 y_{s_1}^3 y_{s_2}^2 y_{s_3} y_{s_4}^2 y_{s_5}^3 + 
 t_1^2 t_2 t_3^2 t_4^2 y_{s_1}^2 y_{s_2}^3 y_{s_3} y_{s_4}^2 y_{s_5}^3 + 
 t_1 t_2^2 t_3^2 t_4^2 y_{s_1}^2 y_{s_2}^3 y_{s_3} y_{s_4}^2 y_{s_5}^3 - 
 t_1^2 t_2^2 t_3^3 t_4^2 y_{s_1}^3 y_{s_2}^2 y_{s_3}^2 y_{s_4}^2 y_{s_5}^3 
 \nn\\
 &&
 - 
 t_1^2 t_2^2 t_3^2 t_4^3 y_{s_1}^3 y_{s_2}^2 y_{s_3}^2 y_{s_4}^2 y_{s_5}^3 - 
 t_1^2 t_2^2 t_3^3 t_4^2 y_{s_1}^2 y_{s_2}^3 y_{s_3}^2 y_{s_4}^2 y_{s_5}^3 - 
 t_1^2 t_2^2 t_3^2 t_4^3 y_{s_1}^2 y_{s_2}^3 y_{s_3}^2 y_{s_4}^2 y_{s_5}^3 - 
 t_1^2 t_2^2 t_3^2 t_4^2 y_{s_1}^3 y_{s_2}^3 y_{s_3} y_{s_4}^3 y_{s_5}^3 
 \nn\\
 &&
 + 
 t_1^3 t_2^3 t_3^3 t_4^3 y_{s_1}^3 y_{s_2}^3 y_{s_3}^3 y_{s_4}^3 y_{s_5}^3
 ~~.
\eea
\vspace{1cm}

\bibliographystyle{JHEP}
\bibliography{mybib}

\providecommand{\href}[2]{#2}\begingroup\raggedright\begin{thebibliography}{10}

\bibitem{Lerche:1989uy}
W.~Lerche, C.~Vafa, and N.~P. Warner, {\it {Chiral Rings in N=2 Superconformal
  Theories}},  {\em Nucl. Phys.} {\bf B324} (1989) 427.

\bibitem{Candelas:1989hd}
P.~Candelas, M.~Lynker, and R.~Schimmrigk, {\it {Calabi-Yau Manifolds in
  Weighted P(4)}},  {\em Nucl. Phys.} {\bf B341} (1990) 383--402.

\bibitem{Greene:1990ud}
B.~R. Greene and M.~R. Plesser, {\it {Duality in Calabi-Yau Moduli Space}},
  {\em Nucl. Phys.} {\bf B338} (1990) 15--37.

\bibitem{morrison-1993-6}
D.~R. Morrison, {\it Mirror symmetry and rational curves on quintic threefolds:
  a guide for mathematicians},  {\em J.AMER.MATH.SOC.} {\bf 6} (1993) 223.

\bibitem{Batyrev:1994hm}
V.~V. Batyrev, {\it {Dual polyhedra and mirror symmetry for Calabi-Yau
  hypersurfaces in toric varieties}},  {\em J. Alg. Geom.} {\bf 3} (1994)
  493--545.

\bibitem{Batyrev:1994ju}
V.~Batyrev and D.~Dais, {\it {Strong McKay correspondence, string theoretic
  Hodge numbers and mirror symmetry}},
  \href{http://xxx.lanl.gov/abs/alg-geom/9410001}{{\tt alg-geom/9410001}}.

\bibitem{Batyrev:1997tv}
V.~V. Batyrev and L.~A. Borisov, {\it {Dual cones and mirror symmetry for
  generalized Calabi-Yau manifolds}}, . In *Greene, B. (ed.): Yau, S.T. (ed.):
  Mirror symmetry II* 71-86.

\bibitem{cox1999mirror}
D.~Cox and S.~Katz, {\em Mirror symmetry and algebraic geometry}.
\newblock Mathematical surveys and monographs. American Mathematical Society,
  1999.

\bibitem{mirrorbook}
K.~Hori, S.~Katz, A.~Klemm, R.~Pandharipande, R.~Thomas, C.~Vafa, R.~Vakil, and
  E.~Zaslow, {\em Mirror symmetry}, vol.~1 of {\em Clay mathematics
  monographs}.
\newblock American Mathematical Society, Providence, RI, 2003.

\bibitem{Feng:2000mi}
B.~Feng, A.~Hanany, and Y.-H. He, {\it {D-brane gauge theories from toric
  singularities and toric duality}},  {\em Nucl. Phys.} {\bf B595} (2001)
  165--200, [\href{http://xxx.lanl.gov/abs/hep-th/0003085}{{\tt
  hep-th/0003085}}].

\bibitem{Feng:2001xr}
B.~Feng, A.~Hanany, and Y.-H. He, {\it {Phase structure of D-brane gauge
  theories and toric duality}},  {\em JHEP} {\bf 08} (2001) 040,
  [\href{http://xxx.lanl.gov/abs/hep-th/0104259}{{\tt hep-th/0104259}}].

\bibitem{Feng:2002zw}
B.~Feng, S.~Franco, A.~Hanany, and Y.-H. He, {\it {Symmetries of toric
  duality}},  {\em JHEP} {\bf 12} (2002) 076,
  [\href{http://xxx.lanl.gov/abs/hep-th/0205144}{{\tt hep-th/0205144}}].

\bibitem{Seiberg:1994pq}
N.~Seiberg, {\it {Electric - magnetic duality in supersymmetric nonAbelian
  gauge theories}},  {\em Nucl.Phys.} {\bf B435} (1995) 129--146,
  [\href{http://xxx.lanl.gov/abs/hep-th/9411149}{{\tt hep-th/9411149}}].

\bibitem{Feng:2001bn}
B.~Feng, A.~Hanany, Y.-H. He, and A.~M. Uranga, {\it {Toric duality as Seiberg
  duality and brane diamonds}},  {\em JHEP} {\bf 12} (2001) 035,
  [\href{http://xxx.lanl.gov/abs/hep-th/0109063}{{\tt hep-th/0109063}}].

\bibitem{2001JHEP...12..001B}
C.~E. {Beasley} and M.~{Ronen Plesser}, {\it {Toric duality is Seiberg
  duality}},  {\em Journal of High Energy Physics} {\bf 12} (Dec., 2001) 1--+,
  [\href{http://xxx.lanl.gov/abs/hep-th/0109053}{{\tt hep-th/0109053}}].

\bibitem{Franco:2003ea}
S.~Franco, A.~Hanany, and Y.-H. He, {\it {A trio of dualities: Walls, trees and
  cascades}},  {\em Fortsch. Phys.} {\bf 52} (2004) 540--547,
  [\href{http://xxx.lanl.gov/abs/hep-th/0312222}{{\tt hep-th/0312222}}].

\bibitem{Hanany:2012hi}
A.~Hanany and R.-K. Seong, {\it {Brane Tilings and Reflexive Polygons}},
  \href{http://xxx.lanl.gov/abs/1201.2614}{{\tt arXiv:1201.2614}}.

\bibitem{Hanany:2005ve}
A.~Hanany and K.~D. Kennaway, {\it {Dimer models and toric diagrams}},
  \href{http://xxx.lanl.gov/abs/hep-th/0503149}{{\tt hep-th/0503149}}.

\bibitem{Franco:2005rj}
S.~Franco, A.~Hanany, K.~D. Kennaway, D.~Vegh, and B.~Wecht, {\it {Brane Dimers
  and Quiver Gauge Theories}},  {\em JHEP} {\bf 01} (2006) 096,
  [\href{http://xxx.lanl.gov/abs/hep-th/0504110}{{\tt hep-th/0504110}}].

\bibitem{Franco:2005sm}
S.~Franco {\em et.~al.}, {\it {Gauge theories from toric geometry and brane
  tilings}},  {\em JHEP} {\bf 01} (2006) 128,
  [\href{http://xxx.lanl.gov/abs/hep-th/0505211}{{\tt hep-th/0505211}}].

\bibitem{Hanany:2005ss}
A.~Hanany and D.~Vegh, {\it {Quivers, tilings, branes and rhombi}},  {\em JHEP}
  {\bf 10} (2007) 029, [\href{http://xxx.lanl.gov/abs/hep-th/0511063}{{\tt
  hep-th/0511063}}].

\bibitem{Hanany:2006nm}
A.~Hanany, C.~P. Herzog, and D.~Vegh, {\it {Brane tilings and exceptional
  collections}},  {\em JHEP} {\bf 07} (2006) 001,
  [\href{http://xxx.lanl.gov/abs/hep-th/0602041}{{\tt hep-th/0602041}}].

\bibitem{Kennaway:2007tq}
K.~D. Kennaway, {\it {Brane Tilings}},  {\em Int. J. Mod. Phys.} {\bf A22}
  (2007) 2977--3038, [\href{http://xxx.lanl.gov/abs/0706.1660}{{\tt
  arXiv:0706.1660}}].

\bibitem{Yamazaki:2008bt}
M.~Yamazaki, {\it {Brane Tilings and Their Applications}},  {\em Fortsch.
  Phys.} {\bf 56} (2008) 555--686,
  [\href{http://xxx.lanl.gov/abs/0803.4474}{{\tt arXiv:0803.4474}}].

\bibitem{1997CMaPh.185..495K}
M.~{Kreuzer} and H.~{Skarke}, {\it {On the Classification of Reflexive
  Polyhedra}},  {\em Communications in Mathematical Physics} {\bf 185} (1997)
  495--508, [\href{http://xxx.lanl.gov/abs/hep-th/9512204}{{\tt
  hep-th/9512204}}].

\bibitem{Kreuzer:1998vb}
M.~Kreuzer and H.~Skarke, {\it {Classification of Reflexive Polyhedra in Three
  Dimensions}},  {\em Adv. Theor. Math. Phys.} {\bf 2} (1998) 847--864,
  [\href{http://xxx.lanl.gov/abs/hep-th/9805190}{{\tt hep-th/9805190}}].

\bibitem{Kreuzer:2000qv}
M.~Kreuzer and H.~Skarke, {\it {Reflexive polyhedra, weights and toric
  Calabi-Yau fibrations}},  {\em Rev. Math. Phys.} {\bf 14} (2002) 343--374,
  [\href{http://xxx.lanl.gov/abs/math/0001106}{{\tt math/0001106}}].

\bibitem{Kreuzer:2000xy}
M.~Kreuzer and H.~Skarke, {\it {Complete classification of reflexive polyhedra
  in four dimensions}},  {\em Adv. Theor. Math. Phys.} {\bf 4} (2002)
  1209--1230, [\href{http://xxx.lanl.gov/abs/hep-th/0002240}{{\tt
  hep-th/0002240}}].

\bibitem{2008arXiv0802.3376B}
V.~{Batyrev} and M.~{Kreuzer}, {\it {Constructing new Calabi-Yau 3-folds and
  their mirrors via conifold transitions}},  {\em ArXiv e-prints} (Feb., 2008)
  [\href{http://xxx.lanl.gov/abs/0802.3376}{{\tt arXiv:0802.3376}}].

\bibitem{2008arXiv0809.4681C}
P.~{Candelas} and R.~{Davies}, {\it {New Calabi-Yau Manifolds with Small Hodge
  Numbers}},  {\em ArXiv e-prints} (Sept., 2008)
  [\href{http://xxx.lanl.gov/abs/0809.4681}{{\tt arXiv:0809.4681}}].

\bibitem{Benvenuti:2004dw}
S.~Benvenuti and A.~Hanany, {\it {New results on superconformal quivers}},
  {\em JHEP} {\bf 0604} (2006) 032,
  [\href{http://xxx.lanl.gov/abs/hep-th/0411262}{{\tt hep-th/0411262}}].

\bibitem{Benvenuti:2005wi}
S.~Benvenuti and A.~Hanany, {\it {Conformal manifolds for the conifold and
  other toric field theories}},  {\em JHEP} {\bf 0508} (2005) 024,
  [\href{http://xxx.lanl.gov/abs/hep-th/0502043}{{\tt hep-th/0502043}}].

\bibitem{HananySeong11b}
{\it work in progress}, .

\bibitem{Feng:2005gw}
B.~Feng, Y.-H. He, K.~D. Kennaway, and C.~Vafa, {\it {Dimer models from mirror
  symmetry and quivering amoebae}},  {\em Adv.Theor.Math.Phys.} {\bf 12} (2008)
  3, [\href{http://xxx.lanl.gov/abs/hep-th/0511287}{{\tt hep-th/0511287}}].

\bibitem{Franco:2011sz}
S.~Franco, {\it {Dimer Models, Integrable Systems and Quantum Teichmuller
  Space}},  {\em JHEP} {\bf 1109} (2011) 057,
  [\href{http://xxx.lanl.gov/abs/1105.1777}{{\tt arXiv:1105.1777}}].

\bibitem{Stienstra:2007dy}
J.~Stienstra, {\it {Hypergeometric Systems in two Variables, Quivers, Dimers
  and Dessins d'Enfants}},  \href{http://xxx.lanl.gov/abs/0711.0464}{{\tt
  arXiv:0711.0464}}.

\bibitem{Butti:2007jv}
A.~Butti, D.~Forcella, A.~Hanany, D.~Vegh, and A.~Zaffaroni, {\it {Counting
  Chiral Operators in Quiver Gauge Theories}},  {\em JHEP} {\bf 11} (2007) 092,
  [\href{http://xxx.lanl.gov/abs/0705.2771}{{\tt arXiv:0705.2771}}].

\bibitem{Franco:2007ii}
S.~Franco, A.~Hanany, D.~Krefl, J.~Park, A.~M. Uranga, {\em et.~al.}, {\it
  {Dimers and orientifolds}},  {\em JHEP} {\bf 0709} (2007) 075,
  [\href{http://xxx.lanl.gov/abs/0707.0298}{{\tt arXiv:0707.0298}}].

\bibitem{Hanany:2008fj}
A.~Hanany, D.~Vegh, and A.~Zaffaroni, {\it {Brane Tilings and M2 Branes}},
  {\em JHEP} {\bf 0903} (2009) 012,
  [\href{http://xxx.lanl.gov/abs/0809.1440}{{\tt arXiv:0809.1440}}].

\bibitem{Forcella:2008bb}
D.~Forcella, A.~Hanany, Y.-H. He, and A.~Zaffaroni, {\it {The Master Space of
  N=1 Gauge Theories}},  {\em JHEP} {\bf 0808} (2008) 012,
  [\href{http://xxx.lanl.gov/abs/0801.1585}{{\tt arXiv:0801.1585}}].

\bibitem{Forcella:2008eh}
D.~Forcella, A.~Hanany, Y.-H. He, and A.~Zaffaroni, {\it {Mastering the Master
  Space}},  {\em Lett.Math.Phys.} {\bf 85} (2008) 163--171,
  [\href{http://xxx.lanl.gov/abs/0801.3477}{{\tt arXiv:0801.3477}}].

\bibitem{2007arXiv0710.1898I}
A.~{Ishii} and K.~{Ueda}, {\it {On moduli spaces of quiver representations
  associated with dimer models}},  {\em ArXiv e-prints} (Oct., 2007)
  [\href{http://xxx.lanl.gov/abs/0710.1898}{{\tt arXiv:0710.1898}}].

\bibitem{Witten:1993yc}
E.~Witten, {\it {Phases of N = 2 theories in two dimensions}},  {\em Nucl.
  Phys.} {\bf B403} (1993) 159--222,
  [\href{http://xxx.lanl.gov/abs/hep-th/9301042}{{\tt hep-th/9301042}}].

\bibitem{Hanany:2010zz}
A.~Hanany and A.~Zaffaroni, {\it {The master space of supersymmetric gauge
  theories}},  {\em Adv.High Energy Phys.} {\bf 2010} (2010) 427891.

\bibitem{Benvenuti:2006qr}
S.~Benvenuti, B.~Feng, A.~Hanany, and Y.-H. He, {\it {Counting BPS operators in
  gauge theories: Quivers, syzygies and plethystics}},  {\em JHEP} {\bf 11}
  (2007) 050, [\href{http://xxx.lanl.gov/abs/hep-th/0608050}{{\tt
  hep-th/0608050}}].

\bibitem{Feng:2007ur}
B.~Feng, A.~Hanany, and Y.-H. He, {\it {Counting Gauge Invariants: the
  Plethystic Program}},  {\em JHEP} {\bf 03} (2007) 090,
  [\href{http://xxx.lanl.gov/abs/hep-th/0701063}{{\tt hep-th/0701063}}].

\bibitem{Hanany:2007zz}
A.~Hanany, {\it {Counting BPS operators in the chiral ring: The plethystic
  story}},  {\em AIP Conf.Proc.} {\bf 939} (2007) 165--175.

\bibitem{Forcella:2008ng}
D.~Forcella, A.~Hanany, and A.~Zaffaroni, {\it {Master Space, Hilbert Series
  and Seiberg Duality}},  {\em JHEP} {\bf 0907} (2009) 018,
  [\href{http://xxx.lanl.gov/abs/0810.4519}{{\tt arXiv:0810.4519}}].

\bibitem{Hori:2000kt}
K.~Hori and C.~Vafa, {\it {Mirror symmetry}},
  \href{http://xxx.lanl.gov/abs/hep-th/0002222}{{\tt hep-th/0002222}}.

\bibitem{Hori:2000ck}
K.~Hori, A.~Iqbal, and C.~Vafa, {\it {D-branes and mirror symmetry}},
  \href{http://xxx.lanl.gov/abs/hep-th/0005247}{{\tt hep-th/0005247}}.

\bibitem{2003math.ph...5057K}
R.~{Kenyon} and J.-M. {Schlenker}, {\it {Rhombic embeddings of planar graphs
  with faces of degree 4}},  {\em ArXiv Mathematical Physics e-prints} (May,
  2003) [\href{http://xxx.lanl.gov/abs/math-ph/0}{{\tt math-ph/0}}].

\bibitem{Aharony:1997bh}
O.~Aharony, A.~Hanany, and B.~Kol, {\it {Webs of (p,q) 5-branes, five
  dimensional field theories and grid diagrams}},  {\em JHEP} {\bf 01} (1998)
  002, [\href{http://xxx.lanl.gov/abs/hep-th/9710116}{{\tt hep-th/9710116}}].

\bibitem{Hanany:2010cx}
A.~Hanany, D.~Orlando, and S.~Reffert, {\it {Sublattice Counting and
  Orbifolds}},  {\em JHEP} {\bf 06} (2010) 051,
  [\href{http://xxx.lanl.gov/abs/1002.2981}{{\tt arXiv:1002.2981}}].

\bibitem{Davey:2011dd}
J.~Davey, A.~Hanany, and R.-K. Seong, {\it {An Introduction to Counting
  Orbifolds}},  {\em Fortsch. Phys.} {\bf 59} (2011) 677--682,
  [\href{http://xxx.lanl.gov/abs/1102.0015}{{\tt arXiv:1102.0015}}].

\bibitem{Hanany:2010ne}
A.~Hanany and R.-K. Seong, {\it {Symmetries of Abelian Orbifolds}},  {\em JHEP}
  {\bf 01} (2011) 027, [\href{http://xxx.lanl.gov/abs/1009.3017}{{\tt
  arXiv:1009.3017}}].

\bibitem{Davey:2010px}
J.~Davey, A.~Hanany, and R.-K. Seong, {\it {Counting Orbifolds}},  {\em JHEP}
  {\bf 06} (2010) 010, [\href{http://xxx.lanl.gov/abs/1002.3609}{{\tt
  arXiv:1002.3609}}].

\bibitem{Hanany:2011iw}
A.~Hanany, V.~Jejjala, S.~Ramgoolam, and R.-K. Seong, {\it {Calabi-Yau
  Orbifolds and Torus Coverings}},  {\em JHEP} {\bf 09} (2011) 116,
  [\href{http://xxx.lanl.gov/abs/1105.3471}{{\tt arXiv:1105.3471}}].

\end{thebibliography}\endgroup


\end{document}